\documentclass[11pt]{article}
\usepackage[T1]{fontenc}
\usepackage[utf8]{inputenc}
\usepackage{lmodern}
\usepackage[margin=1in]{geometry}
\usepackage{microtype}
\usepackage{amsmath,amssymb,amsthm}
\usepackage{mathtools}
\usepackage{booktabs}
\usepackage{array}
\usepackage{float}
\usepackage{algorithm}
\usepackage{algpseudocode}
\usepackage{multicol}
\usepackage{tikz}
\usetikzlibrary{arrows.meta,positioning,calc,fit,backgrounds,shapes.geometric}
\usepackage[breakable]{tcolorbox}
\usepackage[numbers,sort&compress]{natbib}
\usepackage{accsupp}
\usepackage{hyperref}
\usepackage{orcidlink}

\newtheorem{theorem}{Theorem}[section]
\newtheorem{lemma}[theorem]{Lemma}
\newtheorem{proposition}[theorem]{Proposition}
\newtheorem{corollary}[theorem]{Corollary}
\numberwithin{equation}{section}
\theoremstyle{definition}
\newtheorem{definition}[theorem]{Definition}
\newtheorem{example}[theorem]{Example}
\theoremstyle{remark}

\newtcolorbox{notationreminder}{
  breakable, colback=black!3, colframe=black!35, boxrule=0.4pt,
  arc=1.5pt, left=5pt, right=5pt, top=4pt, bottom=4pt
}
\newtcolorbox{proofidea}{
  breakable, colback=blue!2, colframe=blue!35, boxrule=0.4pt,
  arc=1.5pt, left=5pt, right=5pt, top=4pt, bottom=4pt
}

\newcommand{\RDFS}{\ensuremath{\mathsf{RDFS}}}
\newcommand{\pDFS}{p_{\RDFS}}
\newcommand{\Vtree}{V_{\mathrm{tree}}}
\newcommand{\Prob}{\mathbb{P}}
\newcommand{\E}{\mathbb{E}}

\hypersetup{
  pdftitle={Size-Independent Survival Bounds for Routing Without a Map},
  pdfauthor={Kri{\v{s}}j{\=a}nis Petru{\v{c}}e{\c{n}}a},
  pdfkeywords={randomised graph search; unknown topology; depth-first search; bounded degree; survival probability; online algorithms; secret sharing},
  pdfsubject={Theoretical Computer Science: routing in unknown graphs},
  bookmarksnumbered=true,
  bookmarksopen=true,
  bookmarksopenlevel=1,
  colorlinks=true,
  linkcolor=blue!45!black,
  citecolor=blue!45!black,
  urlcolor=blue!45!black
}
\DeclareRobustCommand{\nameSpan}[2]{%
  \BeginAccSupp{method=hex,unicode,space=false,ActualText=#1}#2\EndAccSupp{}}
\tikzset{
  honest/.style={circle, draw=black, thick, fill=blue!8, minimum size=7mm, inner sep=1pt},
  honest small/.style={circle, draw=black, thick, fill=blue!8, minimum size=5.5mm, inner sep=0.5pt},
  cartel/.style={rectangle, draw=black, very thick, dashed, fill=red!12, minimum size=7mm, inner sep=1pt},
  marked leaf/.style={diamond, draw=black, thick, fill=gray!35, minimum size=6mm, inner sep=0.5pt},
  tree edge/.style={thick, black},
  back edge/.style={thick, dashed, black!70},
  fatal edge/.style={thick, dashed, black!70},
  tinylabel/.style={font=\footnotesize}
}

\begin{document}

\title{Size-Independent Survival Bounds for Routing Without a Map}
\author{Kri{\v{s}}j{\=a}nis Petru{\v{c}}e\nameSpan{01460061}{\c{n}a}\\[4pt]
\normalsize Institute of Mathematics and Computer Science,
University of Latvia, Riga, Latvia\\[2pt]
\normalsize \orcidlink{0009-0008-5713-5914}%
  \,\href{https://orcid.org/0009-0008-5713-5914}{0009-0008-5713-5914}}
\date{}

\maketitle

\begin{abstract}
A walker, such as a packet carrying a secret share, must reach a known target in
an unknown graph while avoiding a fixed set of fatal vertices (compromised relays
in the packet interpretation). It learns a vertex's neighbours only on reaching
that vertex safely, remembers its observations, and is promised a safe route.
We prove that a randomised depth-first strategy guarantees a success probability
depending only on the maximum degree and the number of fatal vertices, not on
the size of the graph. Planarity is not required for this guarantee, and planar
examples attain the bound at every sufficiently large graph size. At those sizes,
when the degree cap is at least three, the guarantee is optimal among strict
depth-first strategies. Whether strategies allowed to switch between unfinished
branches can guarantee a larger worst-case survival probability remains open.
For unknown trees and for graphs of maximum degree two, we determine the exact
best worst-case success probability over all routing strategies. Explicit examples compare our strategy with
random-priority frontier routing and ordinary random walks, showing how
exploration order can prevent survival probability from declining as the graph
grows.
\end{abstract}

{\small\noindent\textbf{Keywords.} randomised graph search; unknown topology;
depth-first search; bounded degree; survival probability; online algorithms;
secret sharing.\par}

\begin{multicols}{2}
\tableofcontents
\end{multicols}

\clearpage
\section{Introduction}\label{sec:intro}

A hedgehog is trying to get home through a network of trails hidden by fog. At
each intersection, signs name the neighbouring intersections, but the trails
beyond them remain out of sight. The hedgehog can remember where it has been
and what the signs said. Somewhere in the fog, an owl is waiting.

The owl knows the whole map and knows hedgehogs well: it knows the strategy this
one will use. It chooses where to wait to make the journey as dangerous as
possible, then stays there. It cannot foresee the hedgehog's random choices.
There is one reassurance: at least one route from the starting point to home is
owl-free. How much chance of arriving safely can the hedgehog guarantee without
knowing that route?

The earlier random-priority frontier routing (RPFR) strategy guarantees success
probability at least $1/(n-1)$ on a map with $n$ intersections and one owl
\citep[Theorems~1 and~2]{petrucena2026random}. We give a strategy whose guarantee
instead depends on how many trails meet at an intersection. If this number is at
most $\Delta$, including at the owl's intersection, then
\[
\Pr(\text{arrive safely})\ \ge\
\frac{1}{\Delta(\Delta-1)^{\Delta-1}}\ \ge\ \frac{1}{\Delta^\Delta},
\]
regardless of how large the map is. Our strategy is randomised depth-first search,
abbreviated \RDFS: explore new branches in random order, remember previous
visits, and backtrack when a branch is finished. If a sign names the destination,
go there immediately. Below we give the formal game, allow several owls, and
describe its motivations, especially secret sharing.

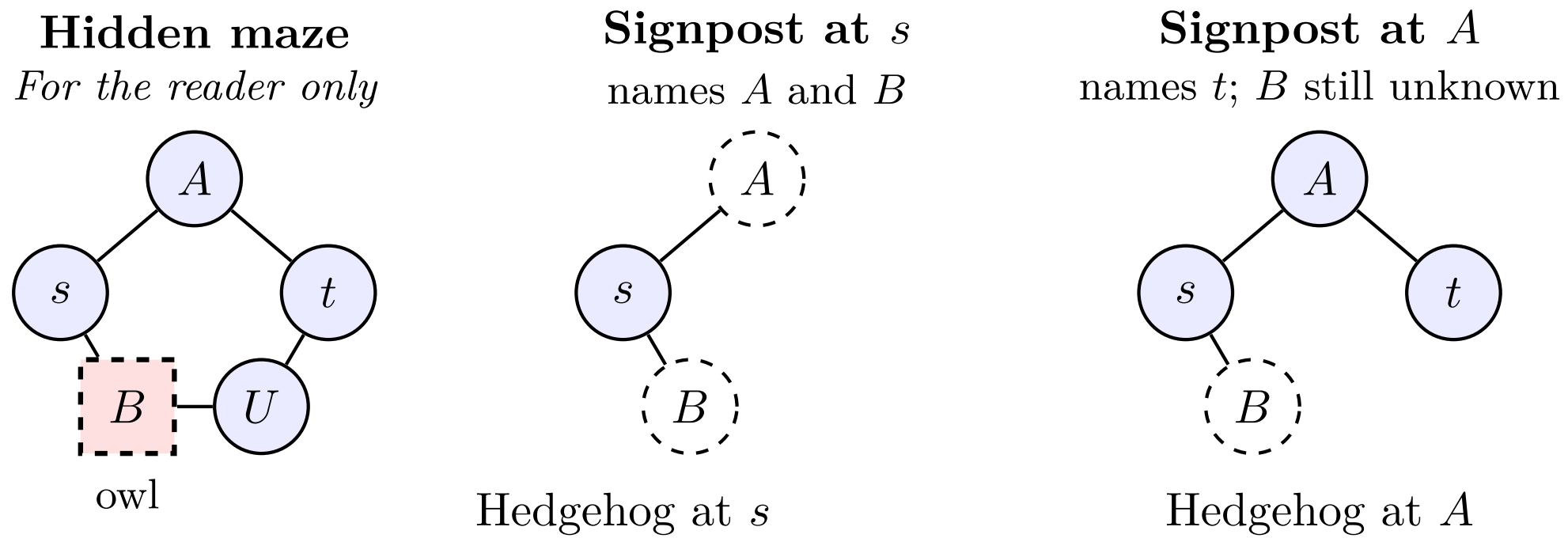
\begin{figure}[H]
\centering
\begin{tikzpicture}[>=Stealth, font=\small, scale=1.12, transform shape,
    unknown/.style={circle, draw=black, dashed, thick, fill=white,
                    minimum size=7mm, inner sep=1pt}]
  \node[honest] (s1) at (0,0) {$s$};
  \node[honest] (a1) at (1,0.85) {$A$};
  \node[honest] (t1) at (2,0) {$t$};
  \node[cartel] (b1) at (0.5,-0.85) {$B$};
  \node[honest] (u1) at (1.5,-0.85) {$U$};
  \draw[tree edge] (s1) -- (a1);
  \draw[tree edge] (a1) -- (t1);
  \draw[tree edge] (t1) -- (u1);
  \draw[tree edge] (u1) -- (b1);
  \draw[tree edge] (b1) -- (s1);
  \node[font=\small\bfseries] at (1,1.95) {Hidden maze};
  \node[font=\footnotesize\itshape] at (1,1.52) {For the reader only};
  \node[tinylabel, below=1pt of b1] {owl};

  \node[honest] (s2) at (4.2,0) {$s$};
  \node[unknown] (a2) at (5.2,0.85) {$A$};
  \node[unknown] (b2) at (4.7,-0.85) {$B$};
  \draw[tree edge] (s2) -- (a2);
  \draw[tree edge] (s2) -- (b2);
  \node[font=\small\bfseries] at (5.2,1.95) {Signpost at $s$};
  \node[font=\footnotesize] at (5.2,1.52) {names $A$ and $B$};
  \node[font=\small] at (4.2,-1.65) {Hedgehog at $s$};

  \node[honest] (s3) at (8.4,0) {$s$};
  \node[honest] (a3) at (9.4,0.85) {$A$};
  \node[honest] (t3) at (10.4,0) {$t$};
  \node[unknown] (b3) at (8.9,-0.85) {$B$};
  \draw[tree edge] (s3) -- (a3);
  \draw[tree edge] (s3) -- (b3);
  \draw[tree edge] (a3) -- (t3);
  \node[font=\small\bfseries] at (9.4,1.95) {Signpost at $A$};
  \node[font=\footnotesize] at (9.4,1.52) {names $t$; $B$ still unknown};
  \node[font=\small] at (9.4,-1.65) {Hedgehog at $A$};
\end{tikzpicture}
\caption{One maze and two signposts, as the hedgehog sees them. The owl at $B$ is
shown only in the reader's view. At $s$ the signpost names the adjacent junctions
$A$ and $B$, but not their status. After a safe arrival at $A$, its signpost names
the destination $t$, while $B$ remains unknown: the hedgehog has not read the
signpost there and does not know what lies beyond it. The name $t$ is known from
the start, and every signpost read is remembered.}
\label{fig:game}
\end{figure}

Figure~\ref{fig:game} shows what the hedgehog can learn from its first few signs.

\subsection{The formal game}\label{sec:intro-game}

We now call the hedgehog a \emph{walker}, the intersections vertices, and the
trails edges. The owl positions are compromised vertices; their set is called the
\emph{cartel}. A \emph{routing policy} is the walker's rule for choosing the next move from
its observations so far and its own random choices.

\begin{samepage}
\begin{definition}[Static unknown-topology instance]\label{def:instance}
An instance is a pair $(G,C)$ with $n$ vertices and $c\ge1$ compromised vertices
(the graph size and owl count). The graph $G$ is connected and undirected, and is
simple: it has no self-loops or parallel edges. Its vertex set is
$\{0,\dots,n-1\}$. The distinct source $s=0$ and target $t=n-1$ are honest;
$C$ is a fixed $c$-element subset of
$V(G)\setminus\{s,t\}$ called the \emph{cartel}, and $G-C$ contains an $s$-$t$
path. The pair is fixed after the policy is chosen and before any of its random
choices is made. The policy knows $n$, $c$, the terminal roles, and any stated
promise about the graph, such as a degree cap. All further information about
the map comes from safe visits, as described below.
\end{definition}
\end{samepage}

Necessarily $n\ge c+2$, since the two honest terminals are outside the cartel.
Such a connected graph has at least three vertices, so its maximum degree is at
least two. Thus a maximum-degree cap $\Delta\ge2$ is automatic; individual
vertices may still have degree one.

At a safe visit to $v$, the signs reveal its neighbour set $N_G(v)$, also called
its \emph{adjacency row}. In particular, $N_G(s)$ is known at the start. Entering
a cartel vertex ends the run in failure before its row is read; reaching $t$
ends it successfully. The walker retains everything it has observed. A
\emph{frontier vertex} is an unvisited vertex named by a visited neighbour.
Since the visited region is connected and safe, the walker can reach any
frontier vertex by first retracing known safe edges. A general policy may
therefore choose any frontier vertex as its next new visit.

For a policy $A$, let $p_A(G,C)$ be the probability that the run reaches $t$ before
any vertex of $C$ is entered. We ask what survival probability one rule can
guarantee across the stated graph class, and how that guarantee depends on $n$ and
$c$.

Memory changes what a rule can do. On the path of $n\ge4$ vertices whose two ends
are the owl and the destination, with the walker starting next to the owl, an
ordinary random walk succeeds with probability $1/(n-1)$, while the depth-first
rule with the destination preference succeeds with probability $1/2$
(Proposition~\ref{prop:walk-comb}). The walk can face the same risky choice
repeatedly; the depth-first rule meets it once and moves on.

\subsection{Motivations for the game}\label{sec:intro-sharing}

Secret sharing is our main motivation for this abstract routing game.
A secret-sharing scheme encodes a key as several shares: enough shares recover
the key, while too few reveal nothing about it. For example, choose a uniformly
random bit string $X_1$, independent of a same-length key $\kappa$, and set
$X_2=\kappa\oplus X_1$. Either share alone reveals nothing about the key, but
$X_1\oplus X_2=\kappa$, where $\oplus$ denotes bitwise exclusive-or (XOR).

Here the walker can be a packet carrying one share, and the owls compromised
relays. Its routing memory travels with it or resides in a controller; the relays
need not remember its past. With confidential links between honest relays, the
routing guarantee bounds one share's chance of arriving unexposed. Turning that
into a key-delivery guarantee also requires enough authentic shares to arrive.
Appendix~\ref{sec:bridge-shares} gives the assumptions, threshold sharing and
privacy amplification in detail.

The same game also captures idealised versions of other tasks. A robot navigating
an unknown environment may be destroyed on entering a hidden hazardous location.
A mobile software agent may expose its state when it enters a compromised
machine. A message may be lost at a relay that silently discards every packet it
receives. These interpretations share the question of reaching a destination
before the first harmful visit. Applying a bound to a particular system requires
the game's assumptions: static hazards, reliable neighbour information, and a
promised safe route.

\subsection{Results in brief}\label{sec:intro-results}

The number $c$ counts compromised vertices. A sharper bound also uses how many
edges lead from the reachable safe region into those vertices. To define this
count, remove the cartel and its incident edges, and let $K$ be the connected
component containing $s$; the safe-path promise puts $t$ there too. Let $d$ count
the edges of the original graph with one endpoint in $K$ and the other in $C$.
One owl may have several such edges: in Figure~\ref{fig:game}, the edges $sB$ and
$UB$ give $d=2$ although $c=1$. For maximum degree at most $\Delta$, \RDFS{}
satisfies
\begin{equation}\label{intro:degree}
\pDFS(G,C)\;\ge\;\frac{1}{\Delta(\Delta-1)^{d-1}}
\;\ge\;\frac{1}{\Delta(\Delta-1)^{\Delta c-1}}
\end{equation}
for every admissible map (Theorem~\ref{thm:degree}). At $\Delta=3$, the second
bound is $(2/3)8^{-c}$, giving $1/12$ for one owl. The rule does not need to know
$d$. Neither bound requires planarity. For $\Delta\ge3$, examples that can be
drawn without edge crossings attain the second bound at every
$n\ge n_0=c(\Delta^2+1)+3$ (Theorem~\ref{thm:sharp}). Thus even restricting the
input to planar graphs cannot improve this rule's worst-case guarantee in that
size range.

We also determine what is possible within \emph{strict depth-first strategies}.
Such a strategy may use arbitrary memory and random choices, but cannot backtrack
past a vertex with an unvisited neighbour or switch to a new branch at an earlier
vertex. Reaching a revealed target is the exception. For the same range
$\Delta\ge3$, $c\ge1$, $n\ge n_0$, no strict depth-first strategy can improve
the guarantee above (Theorem~\ref{thm:strict-dfs}). Whether a strategy allowed to
switch between unfinished branches can guarantee a larger worst-case survival
probability remains open. Corollary~\ref{cor:unrestricted} makes the gap explicit:
for $\Delta\ge3$ and $n\ge2c+2$, the unrestricted value lies between
$W(\Delta,c)=1/[\Delta(\Delta-1)^{\Delta c-1}]$ and $2^{-c}$, also on planar
graphs. For maximum degree two and for trees, we determine the exact best
worst-case probability over all strategies (Proposition~\ref{prop:degree-two}
and Theorem~\ref{thm:trees}).
Section~\ref{sec:comparison} gives explicit comparisons with RPFR and ordinary
random walks; Table~\ref{tab:summary} collects the results.
Section~\ref{comparison:repetitions} converts one-run guarantees into numbers of
independent walkers needed for a prescribed chance that at least one arrives.

Planar graphs form a natural class between trees and unrestricted graphs. They
also provide a useful idealisation for some road and fibre-optic network layouts.
It is therefore natural to ask whether planarity makes routing safer. Our matching
examples show that, in the stated parameter range, it does not improve the
strict-DFS worst-case guarantee.

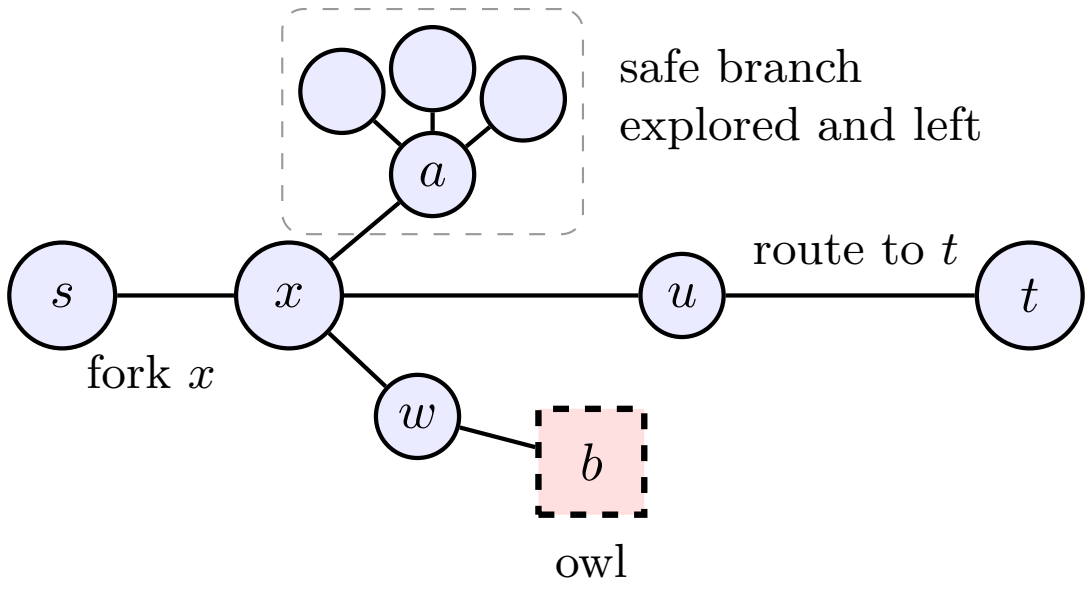
\begin{figure}[H]
\centering
\begin{tikzpicture}[>=Stealth, font=\small]
  \node[honest] (s) at (-1.5,0) {$s$};
  \node[honest] (x) at (0,0) {$x$};
  \node[honest small] (u) at (2.6,0) {$u$};
  \node[honest] (t) at (4.9,0) {$t$};

  \node[honest small] (a) at (0.95,0.8) {$a$};
  \node[honest small] (c1) at (0.35,1.35) {};
  \node[honest small] (c2) at (0.95,1.5) {};
  \node[honest small] (c3) at (1.55,1.3) {};

  \node[honest small] (w) at (0.85,-0.8) {$w$};
  \node[cartel] (b) at (2.0,-1.1) {$b$};

  \draw[tree edge] (s) -- (x);
  \draw[tree edge] (x) -- (u) -- (t);
  \draw[tree edge] (x) -- (a);
  \draw[tree edge] (a) -- (c1);
  \draw[tree edge] (a) -- (c2);
  \draw[tree edge] (a) -- (c3);
  \draw[tree edge] (x) -- (w);
  \draw[tree edge] (w) -- (b);

  \node[draw=black!40, dashed, rounded corners, fit=(a)(c1)(c2)(c3),
        inner sep=3pt] (box) {};
  \node[tinylabel, align=left, anchor=west] at (2.05,1.3)
    {safe branch\\explored and left};

  \node[tinylabel, anchor=east] at (-0.35,-0.5) {fork $x$};
  \node[tinylabel, above=2pt] at (3.75,0) {route to $t$};
  \node[tinylabel, below=1pt of b] {owl};
\end{tikzpicture}
\caption{Depth-first search may explore a whole safe branch and return to $x$
before trying the route through $u$ to $t$. This detour does not cause capture:
success requires choosing $u$ before $w$, but the safe branch at $a$ may come
first. The complete map and safety markings are shown for the reader; the
hedgehog sees only the signs at visited intersections.}
\label{fig:dfs-detour}
\end{figure}

The search need not follow a single safe path: it may explore a large safe
branch and return before continuing towards the target
(Figure~\ref{fig:dfs-detour}). The proof must allow these harmless detours.
It identifies at most $d$ branch choices that matter for survival in an
auxiliary search tree. Since the degree cap also applies to owl vertices,
$d\le\Delta c$, regardless of the size of the rest of the map.

Related work appears near the end, in Section~\ref{sec:related-work}, followed
by the open questions. The appendices contain the full proofs.

\begin{table}[H]
\centering
{\small
\renewcommand{\arraystretch}{1.15}
\begin{tabular}{@{}>{\raggedright\arraybackslash}p{0.32\textwidth}
                >{\raggedright\arraybackslash}p{\dimexpr0.68\textwidth-2\tabcolsep\relax}@{}}
\toprule
Setting & Guarantee or exact value \\
\midrule
Connected simple graphs, max~degree~$\le\Delta$ &
$\pDFS\ge1/[\Delta(\Delta-1)^{\Delta c-1}]$ \\
\addlinespace[0.7em]
Same graphs, strict depth-first strategies; $\Delta\ge3$, $n\ge n_0$ &
$1/[\Delta(\Delta-1)^{\Delta c-1}]$ \\
\addlinespace[0.7em]
Same graphs, all policies; $\Delta\ge3$, $n\ge2c+2$ &
$W(\Delta,c)\le V(n,c,\Delta)\le2^{-c}$; also for the planar value \\
\addlinespace[0.7em]
Maximum degree at most $2$, all policies &
Exact value $1$ at $n=c+2$, and $1/2$ for $n\ge c+3$ \\
\addlinespace[0.7em]
Trees on $n$ vertices &
Exact value in \eqref{tree:answer}; $1$ at $n=c+2$, $2^{-c}$ for $n\ge2c+2$ \\
\addlinespace[0.7em]
Long safe path with one owl, max~degree~$3$ &
RPFR $1/(n-1)$; \RDFS{} $1/4$ \\
\addlinespace[0.7em]
Tree family of Proposition~\ref{prop:walk-comb}, max~degree~$\le3$ &
For fixed $c$, ordinary walk tends to $0$ as $n$ grows; \RDFS{} stays at $2^{-c}$ \\
\bottomrule
\end{tabular}
}
\caption{Summary of survival guarantees and exact values.}
\label{tab:summary}
\medskip
\begin{minipage}{\textwidth}
The first row is a guarantee for \RDFS{}; for $\Delta\ge3$ and $n\ge n_0$ it is
attained by planar examples. In that range, restricting the graph class to planar
graphs leaves both the \RDFS{} worst case and the strict-DFS value unchanged.
The strict-DFS, degree-two and tree rows give exact best worst-case values in
their stated classes. The unrestricted-policy interval is a bound, not an
equality claim. The final two rows compare fixed policies on explicit graph
families.
\end{minipage}
\end{table}

\section{Routing policy and notation}\label{sec:model}

For a stated class $\mathcal{G}(n,c,\Delta)$ of admissible instances with fixed
$n$, $c$ and degree cap $\Delta$, the \emph{best worst-case survival probability} is
\begin{equation}\label{model:value}
V_{\mathcal{G}}(n,c,\Delta)
=\sup_{A}\ \inf_{(G,C)\in\mathcal{G}(n,c,\Delta)} p_A(G,C).
\end{equation}
The policy $A$ is chosen first for the whole class; the instance is then chosen
with full knowledge of $A$. The inner infimum is the worst success probability of
a fixed policy, and it is a minimum here because there are finitely many labelled
instances. The outer supremum is the best guarantee a policy can attain or
approach; a supremum need not be attained by one policy.

We write $V(n,c,\Delta)$ for this value on all admissible instances of maximum
degree at most $\Delta$, and $V_{\mathrm{pl}}(n,c,\Delta)$ for its planar
subclass. Unless a restriction such as strict DFS is stated, the supremum is
over all policies of the model.

If the walker is given the whole graph $G$, it can only do better: it may still
use any rule designed to work without a map. Our guarantees concern the unknown
map, with one rule chosen for the whole stated class.

We allow all randomised rules whose action probabilities are well-defined.
Formally, after each finite observed history, the event of choosing any particular
move must be measurable. Success is a countable union of finite-time arrival
events, so its probability is well-defined. A rule may stop or revisit safe
vertices forever; only arrival at $t$ succeeds. Memory is unlimited.

\begin{samepage}
\subsection{The rule \RDFS}\label{model:rule}

\begin{definition}[Target-first random DFS, \RDFS]\label{def:rdfs}
The rule keeps a stack of visited vertices, which records where to resume after a
branch is exhausted. For each safely visited $v$ it draws an independent uniform
random order $\pi_v$ of $N_G(v)$ when $v$ is first reached safely. It starts with
stack $[s]$. After the row of a newly visited vertex is revealed, the
test for $t$ comes before any scan of its entries. At each step:
\begin{enumerate}
\item If $t$ is a known unvisited neighbour of the visited region, meaning it has
  been named in a revealed row and not entered, take $t$ and succeed. This
  override applies at every step, including between two entries of a row whose
  scan is paused.
\item Otherwise, if the stack top $v$ has an unscanned entry in $\pi_v$, scan
  entries in order. Skip an entry whose label has already been \emph{visited},
  meaning entered by the search; a label merely named in a revealed row is not
  skipped. Take the first unvisited entry $u$, push $u$, and continue from $u$.
\item Pop the top vertex only when its row is exhausted. If the stack becomes
  empty, take and push the least-labelled frontier vertex; if the frontier is
  empty, stop with failure.
\end{enumerate}
Entry into a cartel vertex ends the run in failure before the row is read; entry
into $t$ succeeds and ends the run. The same rule is used on every instance of a
stated class.
\end{definition}
\end{samepage}

On every admissible instance, \RDFS{} ends at $t$ or at a cartel vertex. Each
neighbour list is finite, and each vertex is pushed and popped at most once.
The safe-path promise prevents the search from exhausting its stack before
reaching $t$ unless it has already been captured. Thus the fallback in step~3
is only there to define the rule on other histories. Omitting the instruction
to take a revealed target can only lower the success probability, but the same
degree-based lower bound still holds (Lemma~\ref{lem:plain-bound}). We retain the
override because it is a safe improvement and is used in the exact tree result.

Algorithm~\ref{alg:rdfs} gives the same rule in pseudocode. Here $S$ is the set of
safely visited vertices, $Q$ is the stack, and $R_v$ is the unscanned suffix of
the random order $\pi_v$. The operation $\operatorname{Visit}(u)$ moves to $u$
through known safe vertices and one final edge. Arrival at $t$ ends the run in
success; entry into a cartel vertex ends it in failure before its row is read.
Only a safe nonterminal arrival returns, revealing $N_G(u)$.

\begin{algorithm}[H]
\caption{\RDFS{} as specified in Definition~\ref{def:rdfs}}
\label{alg:rdfs}
\begin{algorithmic}[1]
\State $S\gets\{s\}$; $Q\gets[s]$
\State $R_s\gets$ a uniform random ordering of the known row $N_G(s)$
\Loop
  \State $F\gets\bigl(\bigcup_{v\in S}N_G(v)\bigr)\setminus S$ \Comment{known frontier}
  \If{$t\in F$}
    \State \Call{Visit}{$t$} \Comment{ends the run in success}
  \EndIf
  \If{$Q=[\,]$}
    \If{$F=\varnothing$}
      \State \textbf{halt} with failure
    \EndIf
    \State $u\gets\min F$
  \Else
    \State $v\gets\operatorname{top}(Q)$
    \State Discard leading entries of $R_v$ whose vertices are in $S$
    \If{$R_v=[\,]$}
      \State Pop $Q$; \textbf{continue} the loop
    \EndIf
    \State $u\gets$ remove the first entry of $R_v$
  \EndIf
  \State \Call{Visit}{$u$} \Comment{returns only after a safe nonterminal arrival}
  \State $S\gets S\cup\{u\}$; push $u$ onto $Q$
  \State $R_u\gets$ a fresh independent uniform ordering of the revealed row $N_G(u)$
\EndLoop
\end{algorithmic}
\end{algorithm}

There is also a simple bound on physical movement. On an admissible instance,
implement a push by traversing the first-visit edge and a pop by returning along
that edge. When a newly revealed row first names $t$, take its edge to $t$
immediately. Each first-visit edge is then traversed at most once in each
direction, so a run uses at most
$2(n-1)$ edge traversals. Scanning an already visited neighbour requires no
movement. Thus the survival guarantee is independent of graph size, although
the distance travelled need not be.

\subsection{The hidden boundary and the analysis graph}

Let $K$ be the connected component of $G-C$ containing $s$. Since $G-C$ contains
an $s$-$t$ path, $K$ contains $t$. Every vertex visited before a fatal entry lies
in $K$, and every neighbour of $K$ outside $K$ lies in $C$; otherwise two
components of $G-C$ would be joined. Write $E(G)$ for the edge set of $G$, and
$E(C,K)$ for the subset consisting of edges with one endpoint in $C$ and the
other in $K$. Its number of edges is
\begin{equation}\label{model:d}
d=|E(C,K)|.
\end{equation}
Because $G$ is connected and $C$ is nonempty, $d\ge1$; if the cap applies to every
vertex, including the cartel, then $d\le\Delta c$. The rule is not told $K$ or
$d$; both are analysis parameters.

For the bounded-degree proof, build an analysis graph $H$ from $G[K]$ by adding
one new marked leaf $\ell_{v,b}$ for each edge $vb$ with $v\in K$ and $b\in C$, and
discarding the other components of $G-C$ and the interior of $C$. Then $H$ has $d$
marked leaves and $\deg_H(v)=\deg_G(v)$ for $v\in K$: this is the degree in $G$,
not in $G[K]$. The graph $H$ is an analysis device, never an input to the rule.

\begin{table}[H]
\centering
\small
\begin{tabular}{@{}ll@{}}
\toprule
Symbol & Meaning \\
\midrule
$n,c$ & number of vertices; number of cartel vertices \\
$s=0,\ t=n-1$ & honest source and target \\
$G,\ C$ & hidden connected simple graph; hidden static cartel \\
$A$ & a routing policy, as defined above \\
$\mathcal{D}$ & strict depth-first strategies (Definition~\ref{def:strict-dfs}) \\
$K$ & source component of $G-C$; contains $t$ \\
$E(C,K)$ & edges with one endpoint in $C$ and the other in $K$ \\
$d$ & $|E(C,K)|$, the hidden boundary size; analysis only \\
$\Delta$ & known maximum-degree cap, including cartel vertices \\
$\RDFS,\ \pDFS(G,C)$ & the rule of Definition~\ref{def:rdfs} and its success probability \\
$H$ & $G[K]$ with one marked leaf per $C$-$K$ edge \\
\bottomrule
\end{tabular}
\caption{Core notation. The marked leaves of $H$ are written $\mathcal{L}$; they are not the
cartel. The quantity $h=n-c-2$ counts safe nonterminal vertices.}
\label{tab:notation}
\end{table}

\subsection{Seeded priorities in an implementation}\label{model:seeded}

A practical implementation can recompute scores from a fresh private per-walker
seed $\sigma$ and stable vertex identifiers verified by signatures or
certificates. For \RDFS{}, assign the neighbour $u$ at vertex $v$ the score
$F_\sigma(\operatorname{enc}(\texttt{RDFS},v,u))$, where $F$ is a keyed
pseudorandom function, intended to mimic independent random outputs, and
$\operatorname{enc}$ is an unambiguous encoding. Order neighbours by decreasing
score. A keyed hash such as HMAC
\citep{rfc2104} is a candidate under an appropriate pseudorandom-function
assumption. The current vertex $v$ must be included: hashing only $u$ would
reuse one global ranking and would not supply the independent local orders in
Definition~\ref{def:rdfs}. For RPFR, one global score per $u$ is exactly what is
needed.

Hash the verified identifier, rather than the signature bytes, and keep
identifiers, topology and cartel fixed before choosing the seed. Signatures
authenticate identifiers; they do not by themselves ensure one identifier per
relay or truthful neighbour information. Repetitions use independent seeds, and
any privacy-amplification seed is separate. Finite hash outputs also require a
tie rule, for example ordering equal scores by identifier. The theorems use
exact independent uniform orders. Keyed hashing is a computational implementation
proposal; we do not claim that finite keyed scores preserve the stated guarantees.

\section{The bounded-degree bound}\label{sec:degree}

For every admissible instance of maximum degree at most a fixed $\Delta\ge2$,
one explicit rule, target-first randomised depth-first search (\RDFS), succeeds
with probability at least
\begin{equation}\label{degree:W-def}
W(\Delta,c)=\frac{1}{\Delta(\Delta-1)^{\Delta c-1}},
\end{equation}
independently of $n$. The rule works with or without cycles. The complete proof is in
Appendix~\ref{app:degree-proof}.

\subsection{The guarantee and proof idea}\label{degree:main}

\begin{theorem}[Degree cap]\label{thm:degree}
Fix integers $\Delta\ge2$ and $c\ge1$. Let $G$ be a connected simple undirected
graph on $n\ge c+2$ labelled vertices with maximum degree at most $\Delta$,
source $s=0$ and target $t=n-1$. Let $C$ be a static $c$-subset of
$V(G)\setminus\{s,t\}$ such that $G-C$ contains an $s$-$t$ path, and let $K$ be
the source component of $G-C$. Put $d=|E(C,K)|$. Then $d\ge1$, and
\begin{equation}\label{degree:eq-W}
\pDFS(G,C)\;\ge\;\frac{1}{\Delta(\Delta-1)^{d-1}}
\;\ge\;\frac{1}{\Delta(\Delta-1)^{\Delta c-1}}.
\end{equation}
\end{theorem}

The right-hand side is independent of $n$. At $\Delta=3$ it is $(2/3)8^{-c}$,
equal to $1/12$ at $c=1$; at $\Delta=2$ it is $1/2$. The quantity $d$ is an
analysis parameter, not information available to the rule. The rule does not know
$d$, $K$, $G$, or $C$, and it does not need $\Delta$ as an input.

First ignore the instruction to take a revealed target immediately; this can only
lower the success probability. Replace each edge from the safe component $K$ to
the cartel by a separate marked leaf, as in Section~\ref{sec:model}. This gives
the graph $H$, with marked-leaf set $\mathcal{L}$. Run DFS on $H$ as though all its vertices
were safe, and let $T$ be the resulting search tree. Up to the first marked leaf,
this run makes exactly the same choices as the original search. Thus
\begin{equation}\label{degree:chain}
\pDFS(G,C)
\;\ge\; p_{\mathrm{plain}}(G,C)
\;=\;\Prob_H(t\text{ before }\mathcal{L})
\;=\;\E_T\bigl[\Prob(\mathrm{Success}\mid T)\bigr]
\;\ge\; W(\Delta,c).
\end{equation}
Here $p_{\mathrm{plain}}$ denotes DFS without the target override, and
$\mathrm{Success}$ means reaching $t$ before any marked leaf. The tree $T$ is
random: $\E_T$ averages over all possible DFS trees, weighted by the probabilities
induced by the random neighbour orders. Explicitly,
\[
\E_T\bigl[\Prob(\mathrm{Success}\mid T)\bigr]
=\sum_{\tau:\,\Prob(T=\tau)>0}
\Prob(T=\tau)\,\Prob(\mathrm{Success}\mid T=\tau).
\]
The tree records parent edges but not the order in which its branches are
explored. We show that the conditional success probability is at least
$W(\Delta,c)$ for every possible tree. Their weighted average therefore has the
same lower bound, without needing to compute how likely each tree is.

Figure~\ref{fig:proof-map} shows the two ingredients of this conditional bound.
First, count neighbours, not just tree branches. At $s$, the target branch
contains one neighbour, $v_1$, while the other branch contains two, $u$ and $z$.
The chance of exploring the target branch first, conditional on the depicted
tree, is therefore $1/(1+2)=1/3$. The next two risky vertices, $v_1$ and $v_2$,
each contribute $1/2$. Second, each risky vertex can be assigned a different
marked leaf in a branch away from the target: $s$ uses $\ell_3$, $v_1$ uses
$\ell_1$, and $v_2$ uses $\ell_2$.

\begin{figure}[H]
\centering
\begin{tikzpicture}[>=Stealth, font=\small, x=1cm, y=1cm]
  \node[font=\footnotesize\bfseries] at (2.3,1.25) {Original graph $G$};
  \node[font=\footnotesize\bfseries] at (10.1,1.25) {Analysis graph $H$, DFS tree $T$};

  \coordinate (ls)  at (0,0);
  \coordinate (lv1) at (1.4,0);
  \coordinate (lv2) at (2.8,0);
  \coordinate (lq)  at (4.2,0);
  \coordinate (lt)  at (5.6,0);
  \coordinate (lu)  at (-0.9,-1.0);
  \coordinate (lw)  at (-0.9,-2.2);
  \coordinate (lz)  at (0,-2.2);
  \coordinate (lb)  at (2.1,-2.2);

  \coordinate (rs)  at (7.8,0);
  \coordinate (rv1) at (9.2,0);
  \coordinate (rv2) at (10.6,0);
  \coordinate (rq)  at (12.0,0);
  \coordinate (rt)  at (13.4,0);
  \coordinate (ru)  at (6.9,-1.0);
  \coordinate (rw)  at (6.9,-2.2);
  \coordinate (rz)  at (7.8,-2.2);
  \coordinate (re1) at (9.2,-1.3);
  \coordinate (re2) at (10.6,-1.3);
  \coordinate (re3) at (6.9,-3.4);

  \draw[tree edge] (ls) -- (lv1) -- (lv2) -- (lq) -- (lt);
  \draw[tree edge] (ls) -- (lu) -- (lw) -- (lz) -- (ls);
  \draw[tree edge] (lb) -- (lv1);
  \draw[tree edge] (lb) -- (lv2);
  \draw[tree edge] (lb) .. controls (1.6,-3.4) and (0.2,-3.4) .. (lw);

  \draw[tree edge] (rs) -- (rv1) -- (rv2) -- (rq) -- (rt);
  \draw[tree edge] (rs) -- (ru) -- (rw) -- (rz);
  \draw[tree edge] (rv1) -- (re1);
  \draw[tree edge] (rv2) -- (re2);
  \draw[tree edge] (rw) -- (re3);
  \draw[line width=1.6pt] (rs) -- (rv1) -- (rv2) -- (rq) -- (rt);
  \draw[back edge] (rs) -- (rz);

  \node[honest] (Ls)  at (ls)  {$s$};
  \node[honest] (Lv1) at (lv1) {$v_1$};
  \node[honest] (Lv2) at (lv2) {$v_2$};
  \node[honest] (Lq)  at (lq)  {$q$};
  \node[honest] (Lt)  at (lt)  {$t$};
  \node[honest] (Lu)  at (lu)  {$u$};
  \node[honest] (Lw)  at (lw)  {$w$};
  \node[honest] (Lz)  at (lz)  {$z$};
  \node[cartel] (Lb)  at (lb)  {$b$};

  \node[honest] (Rs)  at (rs)  {$s$};
  \node[honest] (Rv1) at (rv1) {$v_1$};
  \node[honest] (Rv2) at (rv2) {$v_2$};
  \node[honest] (Rq)  at (rq)  {$q$};
  \node[honest] (Rt)  at (rt)  {$t$};
  \node[honest] (Ru)  at (ru)  {$u$};
  \node[honest] (Rw)  at (rw)  {$w$};
  \node[honest] (Rz)  at (rz)  {$z$};
  \node[marked leaf, label={[tinylabel]below:$\ell_1$}] (Re1) at (re1) {};
  \node[marked leaf, label={[tinylabel]below:$\ell_2$}] (Re2) at (re2) {};
  \node[marked leaf, label={[tinylabel]below:$\ell_3$}] (Re3) at (re3) {};

  \draw[thick] (Rs)  circle[radius=0.50cm];
  \draw[thick] (Rv1) circle[radius=0.50cm];
  \draw[thick] (Rv2) circle[radius=0.50cm];
  \node[tinylabel] at (7.8,0.80)  {$1/3$};
  \node[tinylabel] at (9.2,0.80)  {$1/2$};
  \node[tinylabel] at (10.6,0.80) {$1/2$};

  \draw[line width=1.6pt] (4.3,-4.45) -- (4.9,-4.45);
  \node[tinylabel, anchor=west] at (5.0,-4.45) {target path};
  \draw[back edge] (8.0,-4.45) -- (8.6,-4.45);
  \node[tinylabel, anchor=west] at (8.7,-4.45) {non-tree edge};
\end{tikzpicture}
\[
\Prob_H(t\text{ before }\mathcal{L}\mid T)
=\frac13\cdot\frac12\cdot\frac12
=\frac1{12}=W(3,1).
\]
\caption{The proof on an instance with $\Delta=3$, $c=1$ and $d=3$.
Left: one cartel vertex has three contacts with the safe component. Right:
each contact becomes a marked leaf (diamond). Solid edges form one possible
DFS tree $T$, with a bold path to $t$; the dashed edge is outside $T$.
The ringed vertices are risky, with the conditional factors shown. At $s$,
two neighbours enter the same bad subtree, giving $1/3$ rather than $1/2$.
The three risks use three distinct marked leaves.}
\label{fig:proof-map}
\end{figure}

On this $G$, entering the cycle at $s$ forces DFS to visit $b$ before returning.
For \RDFS{}, success therefore requires following the path at $s$, $v_1$ and $v_2$, with
probability $\frac13\cdot\frac12\cdot\frac12=1/12$. The target first becomes
visible at $q$, after all three risks, so the target override gives no advantage
on this example.

In general, group the neighbours below each vertex by the child subtree of $T$
containing them. A group with more neighbours has more chances to be explored
first, so the conditional ordering law weights groups by their sizes. Along the
tree path to $t$, each risky branch point has a marked leaf in a branch away
from that path. These branches are disjoint, so there are at most $d$ such
points. Each contributes a factor at least $1/(\Delta-1)$, except that the root
may contribute $1/\Delta$. The conditional ordering events at distinct vertices
are independent, so multiplying gives the bound. Appendix~\ref{app:degree-proof}
proves this ordering law and the leaf count.

\subsection{A matching example for the fixed rule}\label{degree:sharp}

The bound is attained on examples that can be drawn without edge crossings
(planar graphs). The construction below works at every sufficiently large size.

\begin{theorem}[Matching example]\label{thm:sharp}
Fix $\Delta\ge3$, $c\ge1$, put $\ell_s=\Delta c$ and $n_0=c(\Delta^2+1)+3$. For
every $n\ge n_0$ there is an admissible pair $(G_n,C)$ on $n$ labelled vertices,
with maximum degree at most $\Delta$, $G_n$ planar, $|C|=c$, $G_n-C$ connected,
$d=\Delta c$, and
\begin{equation}\label{degree:eq-sharp}
\pDFS(G_n,C)=\frac{1}{\Delta(\Delta-1)^{\Delta c-1}}=W(\Delta,c).
\end{equation}
Consequently, for each fixed $n\ge n_0$, the infimum of $\pDFS(G,C)$ over
admissible instances on $n$ vertices with maximum degree at most $\Delta$ and
$|C|=c$ equals this value, and the same infimum is attained on the planar
subclass.
\end{theorem}

The threshold $n_0$ is sufficient, not claimed minimal:
Figure~\ref{fig:proof-map} already attains $W(3,1)$ for \RDFS{} with nine
vertices, below $n_0=13$ for those parameters.

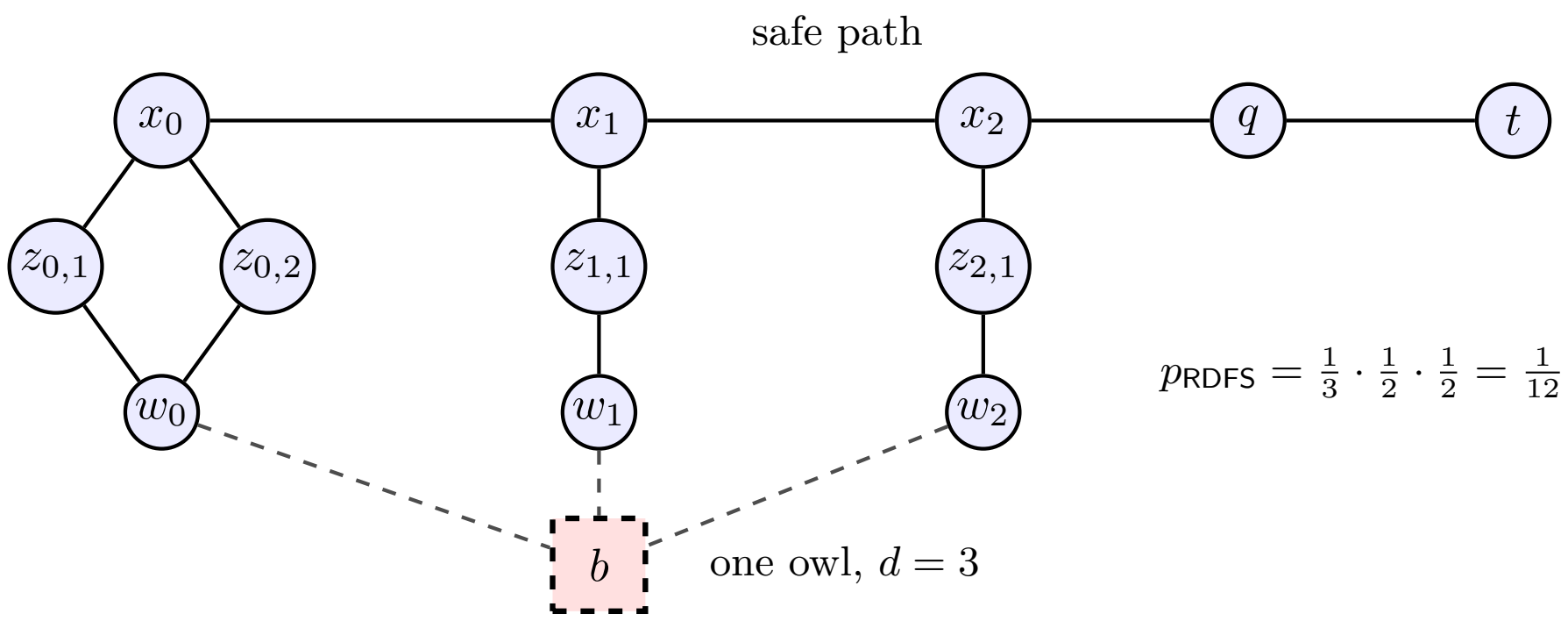
\begin{figure}[H]
\centering
\begin{tikzpicture}[>=Stealth, font=\small]
  \node[honest] (x0) at (0,0) {$x_0$};
  \node[honest] (x1) at (3.3,0) {$x_1$};
  \node[honest] (x2) at (6.2,0) {$x_2$};
  \node[honest small] (q) at (8.2,0) {$q$};
  \node[honest small] (t) at (10.2,0) {$t$};
  \draw[tree edge] (x0) -- (x1);
  \draw[tree edge] (x1) -- (x2);
  \draw[tree edge] (x2) -- (q);
  \draw[tree edge] (q) -- (t);

  \node[honest] (z01) at (-0.8,-1.1) {$z_{0,1}$};
  \node[honest] (z02) at (0.8,-1.1) {$z_{0,2}$};
  \node[honest] (z11) at (3.3,-1.1) {$z_{1,1}$};
  \node[honest] (z21) at (6.2,-1.1) {$z_{2,1}$};
  \node[honest small] (w0) at (0,-2.2) {$w_0$};
  \node[honest small] (w1) at (3.3,-2.2) {$w_1$};
  \node[honest small] (w2) at (6.2,-2.2) {$w_2$};
  \draw[tree edge] (x0) -- (z01); \draw[tree edge] (z01) -- (w0);
  \draw[tree edge] (x0) -- (z02); \draw[tree edge] (z02) -- (w0);
  \draw[tree edge] (x1) -- (z11); \draw[tree edge] (z11) -- (w1);
  \draw[tree edge] (x2) -- (z21); \draw[tree edge] (z21) -- (w2);

  \node[cartel] (b) at (3.3,-3.35) {$b$};
  \draw[fatal edge] (w0) -- (b);
  \draw[fatal edge] (w1) -- (b);
  \draw[fatal edge] (w2) -- (b);
  \node[tinylabel, anchor=west] at (4.0,-3.35) {one owl, $d=3$};

  \node[tinylabel] at (5.1,0.65) {safe path};
  \node[tinylabel, align=left, anchor=west] at (7.4,-1.9)
    {$\pDFS=\frac13\cdot\frac12\cdot\frac12=\frac1{12}$};
\end{tikzpicture}
\caption{An example attaining the bound at $\Delta=3$, $c=1$: $13$ vertices,
$15$ edges and one owl at $b$. To survive, \RDFS{} must choose $x_1$ at $x_0$,
$x_2$ at $x_1$, and $q$ at $x_2$ before taking any arm towards $b$. The
probabilities are $1/3$, $1/2$ and $1/2$, giving $1/12$. The extra safe vertex
$q$ keeps $t$ out of sight at the last risky choice. The full map is shown for
the reader.}
\label{fig:degree-witness}
\end{figure}

\begin{proof}[Construction and probability]
Take distinct vertices $x_0,\dots,x_{\ell_s-1}$, $q$, $t$, $b_0,\dots,b_{c-1}$,
$w_0,\dots,w_{\ell_s-1}$, and $z_{i,j}$ for $0\le i<\ell_s$, $1\le j\le m_i$, where
$m_0=\Delta-1$ and $m_i=\Delta-2$ for $i\ge1$. The extra safe vertex $q$ keeps
the target out of sight until the search has passed every $x_i$. The edges are
\[
x_ix_{i+1}\ (i<\ell_s-1),\quad x_{\ell_s-1}q,\quad qt,\quad
w_ib_{\lfloor i/\Delta\rfloor},\quad
x_iz_{i,j},\quad z_{i,j}w_i.
\]
Set $s=x_0$, target $t$, and $C=\{b_0,\dots,b_{c-1}\}$. The families are disjoint
and the graph is simple. Its vertex count is
$n_0=\Delta\ell_s+c+3=c(\Delta^2+1)+3$. Each $x_i$ has degree
$\Delta$; $q,t,z_{i,j}$ have degrees $2,1,2$; $w_0$ has degree $\Delta$ and
$w_i$ for $i>0$ degree $\Delta-1$; each $b_l$ is adjacent to the $\Delta$ hubs
with $\lfloor i/\Delta\rfloor=l$, so has degree $\Delta$. Thus the maximum degree
is $\Delta$, $G-C$ is the path $x_0,\dots,x_{\ell_s-1},q,t$ with the arms
$x_iz_{i,j}w_i$ attached, hence
connected, and $d=\Delta c=\ell_s$.

The target is revealed only at $q$. If the search leaves the path at $x_i$, it
reaches $w_i$. Each remaining arm reaches another $z_{i,j}$ whose two neighbours
are already visited, so the search returns to $w_i$. It therefore enters
$b_{\lfloor i/\Delta\rfloor}$ before it can return to $x_i$, and fails.
Success requires the forward neighbour first at every $x_i$;
these choices are independent, with probability $1/\Delta$ at $x_0$ and
$1/(\Delta-1)$ elsewhere, so
\[
\pDFS(G_{n_0},C)
=\frac{1}{\Delta}\Bigl(\frac{1}{\Delta-1}\Bigr)^{\ell_s-1}
=\frac{1}{\Delta(\Delta-1)^{\Delta c-1}}.
\]
When $q$ is reached its row shows $t$, and the target is taken.

Appendix~\ref{app:degree-proof} verifies planarity and extends the construction
to every $n\ge n_0$ by lengthening the final safe path after $q$. These extra vertices
do not change the success probability.
Figure~\ref{fig:degree-witness} shows the example at $\Delta=3$, $c=1$.
\end{proof}

\subsection{Optimality within strict DFS}\label{degree:strict}

The matching value is also the best worst-case guarantee over a natural strict
depth-first class, even when neighbour choices use the whole observed history.

The fresh arms matter for this stronger claim. In Figure~\ref{fig:proof-map},
$b$ has already been named when $v_2$ is reached, while $q$ is new; a rule can
favour the new label without knowing which vertex is fatal. In the matching
construction, the first visit to each $x_i$ along a surviving run instead offers
fresh forward and arm labels, whose roles the observed history cannot
distinguish under the uniform relabelling used in the upper-bound proof.

\begin{definition}[Strict depth-first policies]\label{def:strict-dfs}
A policy in $\mathcal{D}$ keeps a stack of distinct safely visited vertices,
initially $[s]$, with $N(s)$ revealed at initialisation. A \emph{frame} is one
entry of this stack; the active frame is its top entry. The following rules
define the class.
\begin{enumerate}
\item The policy may take $t$ whenever it is a known frontier vertex. Apart from
  this target exception, every next \emph{first} visit must be to an unvisited
  neighbour of the stack top; a safe nonterminal arrival pushes the new vertex.
\item Only the top frame may be popped, and only when its vertex has no unvisited
  neighbour.
  While such a neighbour remains, the frame cannot be abandoned, suspended in
  favour of an ancestor's new branch, or bypassed by a first visit from another
  frontier vertex.
\item Known safe vertices may be revisited freely. These moves neither reveal a
  new row nor change the active frame, so they do not permit a new first visit
  outside the rule in item~1.
\item Choices may use the whole observed history, labels, $n,c,\Delta$, unlimited
  memory and internal randomness. They need not be uniform or independent, but
  must be measurable as in Section~\ref{sec:model}.
\end{enumerate}
An empty stack, a voluntary stop, or an endless run is failure. No row or degree
of an unvisited vertex and no physical port order is observed; a label appearing
in a row is not thereby visited. The rule \RDFS{} belongs to $\mathcal{D}$ on
every admissible instance.
\end{definition}

\begin{samepage}
\begin{theorem}[Optimality within strict DFS]\label{thm:strict-dfs}
Fix $\Delta\ge3$, $c\ge1$, and $n\ge n_0=c(\Delta^2+1)+3$. Let
$\mathcal{I}(n,c,\Delta)$ be the admissible instances of Theorem~\ref{thm:degree},
and let $\mathcal{I}_{\mathrm{pl}}$ be its planar subclass. Then
\begin{equation}\label{degree:eq-strict}
\sup_{A\in\mathcal{D}}\ \inf_{(G,C)\in\mathcal{I}(n,c,\Delta)} p_A(G,C)
= W(\Delta,c)
= \frac{1}{\Delta(\Delta-1)^{\Delta c-1}},
\end{equation}
and the same equality holds with $\mathcal{I}_{\mathrm{pl}}$ in place of
$\mathcal{I}$.
\end{theorem}
\end{samepage}

\noindent The lower bound is \RDFS{} from Theorem~\ref{thm:degree}; the upper bound
uses the planar matching example of Theorem~\ref{thm:sharp} and relabels its
non-terminal vertices uniformly at random, fixing source label $0$ and target
label $n-1$. On first reaching $x_i$, the walker sees $k_0=\Delta$ new neighbour
labels at the root and $k_i=\Delta-1$ elsewhere. Its observations do not
distinguish which new label continues towards $t$: each is equally likely to do
so. Even a rule that prefers particular labels therefore chooses the continuation
with conditional probability at most $1/k_i$. Every other choice enters an arm
from which strict DFS cannot return before capture. Multiplying the conditional
bounds gives $W(\Delta,c)$; independence between levels is not needed.
Appendix~\ref{app:strict-proof} gives the full argument, including rules that
stop or revisit safe vertices forever.

The restriction on unfinished branches is essential to this proof. Every arm
vertex $z_{i,j}$ is safe. An unrestricted policy may enter it, read its row, and
return to $x_i$ without continuing to $w_i$ and the cartel. Strict DFS instead
keeps that frame active while $w_i$ is unvisited. Thus choosing an arm forces
failure within $\mathcal{D}$, but the same implication does not hold for an
unrestricted policy. The theorem establishes optimality within the stated
class, not among all routing policies.

\subsection{The degree-two case}\label{degree:two}

\begin{proposition}[Exact degree-two value]\label{prop:degree-two}
Fix $c\ge1$ and $n\ge c+2$. For graphs of maximum degree at most $2$,
\[
V(n,c,2)=V_{\mathrm{pl}}(n,c,2)=
\begin{cases}
1, & n=c+2,\\
1/2, & n\ge c+3.
\end{cases}
\]
The rule \RDFS{} attains this best worst-case guarantee over all policies.
\end{proposition}

\begin{proof}
For $n\ge c+3$, the lower bound \eqref{degree:eq-W} gives $1/2$ on every
admissible instance. For the upper bound put $h=n-c-2\ge1$ and use the path
\[
b_c-\cdots-b_1-s-q_1-\cdots-q_h-t,
\qquad C=\{b_1,\dots,b_c\}.
\]
Fix the terminal labels and uniformly relabel all nonterminals. Conditional on
the revealed source row, its two neighbour labels are equally likely to be
$b_1$ and $q_1$. For any deterministic policy, the first move therefore enters
the safe branch with probability at most $1/2$; stopping instead cannot
succeed. Choosing $b_1$ fails immediately. Averaging over a randomised policy's
internal choices preserves the upper bound, so some fixed labelled path has
success probability at most $1/2$. The path is planar, and \RDFS{} succeeds
on it exactly when its first choice is $q_1$.

At $n=c+2$, removing the cartel leaves only $s$ and $t$. Admissibility forces
the edge $st$, which \RDFS{} takes immediately, giving value $1$.
\end{proof}

\section{The exact value on trees}\label{sec:trees}

Suppose the walker is told that the unknown graph is a tree. There is then a
unique $s$-$t$ path, and the safe-path promise makes that path safe. A walker with
a map could simply follow it and succeed with probability $1$. Without a map,
even unlimited memory cannot remove the uncertainty at each fork. We determine
the best guarantee exactly; Appendix~\ref{app:tree-proof} supplies the full proof.

Fix $c\ge1$, $n\ge c+2$, and put $h=n-c-2$. Thus $h$ is the number of vertices
that are neither an endpoint nor in the cartel; call these the safe nonterminal
vertices. For $h>0$ write $c=\alpha h+\beta$ with $0\le \beta<h$, so
$\alpha=\lfloor c/h\rfloor$.

\begin{theorem}[Exact tree value]\label{thm:trees}
Fix $c\ge1$ and $n\ge c+2$, and let $h=n-c-2$. Let
$\mathcal{I}_{\mathrm{tree}}(n,c)$ be the admissible pairs $(G,C)$ in which $G$
is a tree on $\{0,\dots,n-1\}$ with $s=0$ and $t=n-1$. Define
\begin{equation}\label{tree:value}
\Vtree(n,c)=\sup_{A}\ \min_{(G,C)\in\mathcal{I}_{\mathrm{tree}}(n,c)}
p_A(G,C),
\end{equation}
the supremum over all rules of the model, each knowing $n$, $c$, and the tree
promise only. Then
\begin{equation}\label{tree:answer}
\Vtree(n,c)=
\begin{cases}
1, & h=0,\\[2pt]
\dfrac{1}{(\alpha+1)^{h-\beta}(\alpha+2)^{\beta}}, & h>0,
\end{cases}
\end{equation}
with $\alpha,\beta$ as above; write $\psi(n,c)$ for the right-hand side.
\end{theorem}

Once $n\ge2c+2$, the answer stays at $2^{-c}$, one half multiplied by itself
$c$ times. At the smallest possible size, $n=c+2$, the answer is $1$.

\subsection{Lower bound: a target-first DFS product}\label{tree:lower}

Fix an admissible tree instance. Let $P=(v_0,\dots,v_L)$ be the unique $s$-$t$
path, with $v_0=s$, $v_L=t$, and $L\ge1$. Root $G$ at $s$ and let
$\mathrm{ch}(v)$ be the children of $v$. Every $s$-$t$ path in a tree equals $P$,
so $P$ is disjoint from $C$, all $c$ cartel vertices lie among the $n-(L+1)$
vertices outside $P$, and $L-1\le h$.

For $i=0,\dots,L-2$ define
\begin{equation}\label{tree:Di}
D_i=\{\,w\in\mathrm{ch}(v_i): w\ne v_{i+1}
\text{ and the subtree at }w\text{ contains a cartel vertex}\,\},
\qquad d_i=|D_i|.
\end{equation}
The subtrees rooted at the vertices of $D_0\cup\dots\cup D_{L-2}$ are pairwise
disjoint and each contains a cartel vertex, so $d_0+\dots+d_{L-2}\le c$. The index
range stops at $L-2$: $v_{L-1}$ is the parent of $t$, and a target-first rule
takes $t$ as soon as $v_{L-1}$ is visited, before any child outside the path is probed. On
a tree the stored order of the children is a uniform permutation.

\begin{lemma}[DFS lower bound]\label{lem:dfs-lower}
For every $(G,C)\in\mathcal{I}_{\mathrm{tree}}(n,c)$,
\begin{equation}\label{tree:dfsvalue}
\pDFS(G,C)=\prod_{i=0}^{L-2}\frac{1}{d_i+1}\;\ge\;\psi(n,c).
\end{equation}
\end{lemma}

\noindent Safe subtrees may be explored first and left without harm. Among the
$d_i+1$ children $\{v_{i+1}\}\cup D_i$, however, the continuation $v_{i+1}$ must
come first: exploring any of the others reaches the cartel before returning.
These children are equally likely to be first, independently at each $v_i$.
Hence success is exactly the intersection of these ordering events and
the product formula holds. If $h=0$ then $L=1$ by $L-1\le h$, the empty product
equals $1=\psi(n,c)$, and the bound holds. Assume $h\ge1$; padding the list with
$h-(L-1)$ zeros and applying Lemma~\ref{lem:balanced} gives the bound. The full
proof is in Appendix~\ref{app:tree-proof}.

\begin{lemma}[Balanced product]\label{lem:balanced}
Let $h\ge1$ and $c=\alpha h+\beta$ with $0\le \beta<h$. For any nonnegative integers
$x_1,\dots,x_h$ with $x_1+\dots+x_h\le c$,
\begin{equation}\label{tree:balancedineq}
\prod_{i=1}^{h}(x_i+1)\le(\alpha+1)^{h-\beta}(\alpha+2)^{\beta}.
\end{equation}
\end{lemma}

Applying Lemma~\ref{lem:balanced} to the padded list and taking reciprocals in
\eqref{tree:dfsvalue} proves $\pDFS(G,C)\ge\psi(n,c)$ on every tree instance, so
$\Vtree(n,c)\ge\psi(n,c)$.

\subsection{Upper bound: a randomly labelled comb}\label{tree:upper}

\begin{figure}[H]
\centering
\begin{tikzpicture}[>=Stealth, font=\small]
  \node[honest] (s)  at (0,0)   {$s$};
  \node[honest] (a1) at (2.2,0) {$a_1$};
  \node[honest] (a2) at (4.4,0) {$a_2$};
  \node[honest] (t)  at (6.6,0) {$t$};
  \node[cartel] (z1) at (0,1.7)   {$z_1$};
  \node[cartel] (z2) at (0,-1.7)  {$z_2$};
  \node[cartel] (z3) at (2.2,1.7) {$z_3$};

  \draw[tree edge] (s) -- (z1);
  \draw[tree edge] (s) -- (z2);
  \draw[tree edge] (a1) -- (z3);

  \draw[line width=1.6pt] (s) -- (a1) -- (a2) -- (t);

  \node[tinylabel] at (0.42,-0.5) {$1/3$};
  \node[tinylabel] at (2.2,-0.5)  {$1/2$};

  \node[tinylabel] at (5.2,0.32) {spine};
  \node[tinylabel, anchor=east] at (-0.55,1.7) {teeth};
\end{tikzpicture}
\caption{A comb is a path (the spine) with pendant leaves (the teeth). Here
$n=7$, $c=3$: the safe spine is $s,a_1,a_2,t$, and the teeth $z_1,z_2,z_3$ are
compromised. The first two choices have survival factors $1/3$ and $1/2$,
giving $1/6$. Randomly choosing the safe label in each layer makes the average
success of any policy at most $1/6$.}
\label{fig:tree-comb}
\end{figure}
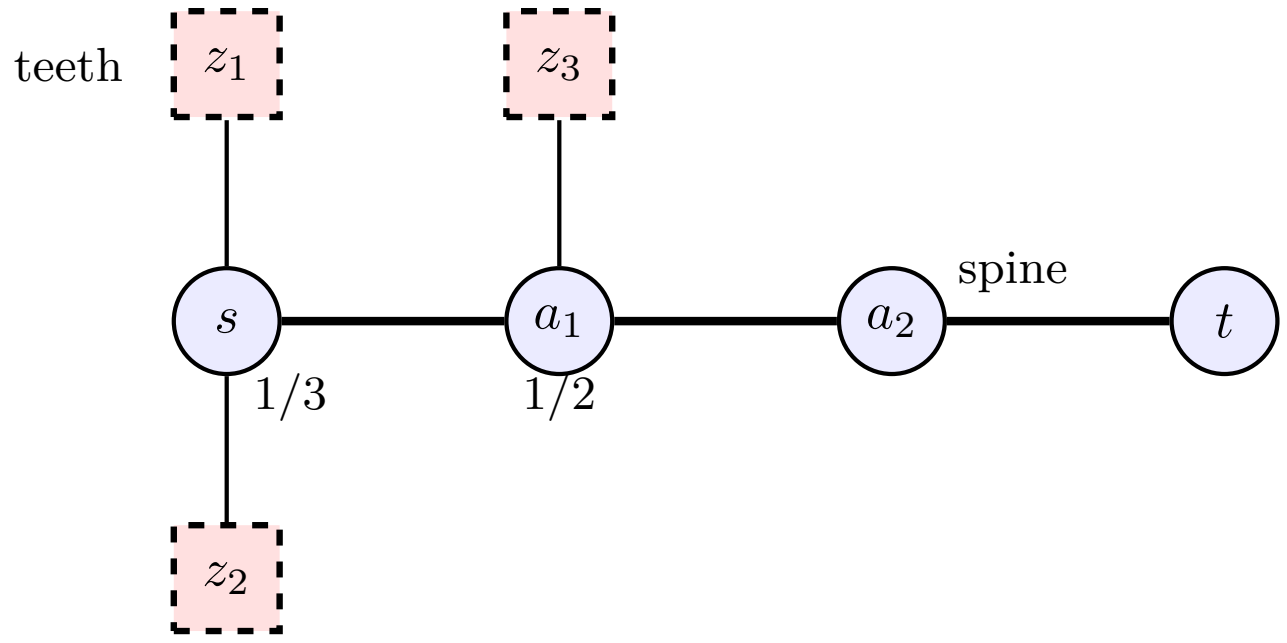

Assume $h\ge1$. Partition the $n-2$ non-endpoint labels into layers
$\Lambda_1,\dots,\Lambda_h$ of sizes $\alpha+2$ for $1\le i\le \beta$ and
$\alpha+1$ for $\beta<i\le h$; it exists because
$\beta(\alpha+2)+(h-\beta)(\alpha+1)=c+h=n-2$. Independently for each
$i$, sample a uniform $a_i\in\Lambda_i$, let $C$ be the union of the labels
$\Lambda_i\setminus\{a_i\}$. Join $s,a_1,\dots,a_h,t$ in a path, and attach the
other labels in layer $i$ as leaves at the preceding path vertex. The path is
the \emph{spine} of the comb, and the attached leaves are its \emph{teeth};
Figure~\ref{fig:tree-comb} shows a small example. The result is a tree whose
unique $s$-$t$ path is safe, so every outcome is admissible. Let
$\mu_{\mathrm{tree}}$ be this distribution, sampled once before any internal
random choice of the rule.

\begin{proposition}[Comb upper bound]\label{prop:comb}
For every rule $A$ of the model, there is an instance in the support of
$\mu_{\mathrm{tree}}$ with $p_A(G,C)\le\psi(n,c)$. Consequently
$\Vtree(n,c)\le\psi(n,c)$ for $h\ge1$.
\end{proposition}

\noindent Fix the rule's random choices. Before entering a layer, the walker
knows its labels but cannot tell which one continues the safe path. Its chance
of choosing that label is at most $1/|\Lambda_i|$. Multiplying these conditional
bounds, then averaging over the rule's random choices, gives
$\E_{\mu}p_A\le\prod_i1/|\Lambda_i|=\psi(n,c)$. At least one fixed instance has
success probability no larger than this average. This proves the upper bound;
Appendix~\ref{app:tree-proof} gives the details for arbitrary observed histories.
Together with Lemma~\ref{lem:dfs-lower}, it gives the exact value.

\subsection{Boundary cases and an example}\label{tree:h0}

If $h=0$, then $n=c+2$ and the non-endpoint set has exactly $c$ labels, so
$C=\{1,\dots,n-2\}$ for every admissible instance. In $G-C$ only $s$ and $t$
remain, so reachability forces the edge $st$; after the row of $s$ is read, $t$ is
in the frontier and the target-first rule takes it immediately, so the value is
$1$.

If $h\ge c$, then $\alpha\in\{0,1\}$. If $h=c$, then $\alpha=1$ and $\beta=0$,
so $\psi=(\alpha+1)^{-h}=2^{-c}$. If $h>c$, then $\alpha=0$ and $\beta=c<h$, so
$\psi=1^{h-c}2^{-c}=2^{-c}$. Since $h\ge c$ is equivalent to $n\ge2c+2$,
\begin{equation}\label{tree:plateau}
\Vtree(n,c)=2^{-c}\qquad\text{for }n\ge2c+2.
\end{equation}
At the other end, $\Vtree(c+2,c)=1$.

Take $n=7$ and $c=3$. Then $h=2$ and $3=1\cdot2+1$, so $\alpha=\beta=1$ and
\begin{equation}\label{tree:examplevalue}
\psi(7,3)=\frac{1}{(1+1)^{1}(1+2)^{1}}=\frac{1}{6}.
\end{equation}
The comb prior has layers of sizes $3$ and $2$; Figure~\ref{fig:tree-comb} shows
one outcome. The honest path is $s,a_1,a_2,t$, so $L=3$. The bad children are
$D_0=\{z_1,z_2\}$ and $D_1=\{z_3\}$, while $D_2$ is not formed: $a_2$ is the
penultimate path vertex, and target-first takes $t$ before any child of $a_2$ is
probed. Thus $d_0=2$, $d_1=1$, and the DFS succeeds with probability
$\frac{1}{3}\cdot\frac{1}{2}=\frac16$. The last factor comes from $a_1$:
once $a_2$ is reached, the target is visible. The distribution chooses $a_1$ uniformly in a layer of size $3$ and
$a_2$ uniformly in a layer of size $2$, so every rule succeeds with probability at
most $1/6$ under the prior; the bounds coincide.

\subsection{A consequence for general bounded-degree graphs}\label{tree:general}

The same comb makes the gap for unrestricted policies explicit. Unlike the
matching example in Section~\ref{degree:strict}, its wrong choices enter cartel
vertices immediately, leaving no opportunity to retreat after a safe probe.

\begin{corollary}[Bounds for unrestricted policies]\label{cor:unrestricted}
For $\Delta\ge3$, $c\ge1$ and $n\ge2c+2$,
\begin{equation}\label{eq:unrestricted-interval}
\frac{1}{\Delta(\Delta-1)^{\Delta c-1}}
\le V(n,c,\Delta)
\le V_{\mathrm{pl}}(n,c,\Delta)
\le 2^{-c}.
\end{equation}
\end{corollary}

\begin{proof}
Theorem~\ref{thm:degree} gives the lower bound. Restricting the adversary to
planar graphs cannot decrease the best guarantee, giving the middle inequality.
For the upper bound, $h=n-c-2\ge c$, so the comb distribution in
Section~\ref{tree:upper} has exactly $c$ layers of size two and all other layers
of size one. Every outcome is a tree of maximum degree at most three, hence
belongs to both graph classes. The proof of Proposition~\ref{prop:comb} applies
to every policy on this distribution: before a first visit, every choice of the
safe label gives the same revealed rows and satisfies both promises. No degree
of an unvisited vertex is observed, so that label remains uniform. The average
success probability is therefore at most $2^{-c}$, and some fixed instance has
no larger success probability.
\end{proof}

At degree cap three, the bounds proved here give
$(2/3)8^{-c}\le V(n,c,3)\le V_{\mathrm{pl}}(n,c,3)\le2^{-c}$.
The size condition $n\ge2c+2$ is weaker than the condition
$n\ge c(\Delta^2+1)+3$ for strict-DFS optimality. The upper bound applies to
all policies; it does not establish that \RDFS{} is optimal outside strict DFS.

\section{Why the exploration order matters}\label{sec:comparison}

Different exploration orders can give very different chances of survival on the
same map. We first compare \RDFS{} with random-priority frontier routing (RPFR), then with the
ordinary random walk.

For each $n\ge5$, let $G_n$ have distinct vertices $x_0=s$,
$x_1,\dots,x_{n-2}=t$, and $b$, with $C=\{b\}$. Its edges are the path edges
$x_ix_{i+1}$ for $0\le i<n-2$, together with $sb$ and $x_1b$. The graph is simple,
connected, and has maximum degree $3$; entry into $b$ is fatal before its row is
read, and $G_n-C$ is the honest path.
(i) \emph{RPFR} assigns independent identically distributed continuous
scores to non-source labels (equivalently, a uniform random priority permutation)
and visits the highest-scored frontier vertex, with no target preference.
(ii) \emph{Target-first RPFR} also takes $t$ immediately whenever it is
in the frontier. (iii) \RDFS{} is the rule of Definition~\ref{def:rdfs}. Scores may
be sampled at the start or when their labels first appear; the distribution is the same.

\begin{example}[Priority on a long safe path]\label{ex:priority-tail}
For every $n\ge5$, on $G_n$ the three success probabilities are
\begin{equation}\label{ex:priority-values}
\begin{array}{ll}
\text{RPFR:} & 1/(n-1),\\[2pt]
\text{target-first RPFR:} & 1/(n-2),\\[2pt]
\text{\RDFS:} & 1/4.
\end{array}
\end{equation}
\end{example}

\begin{proof}
Before a fatal visit or success the visited set is an honest path prefix.
Initially the frontier is $\{b,x_1\}$; each safe extension reveals only the next
path vertex, while $b$ remains unvisited. Thus at $x_i$, for $0\le i\le n-3$, the
frontier is exactly $\{b,x_{i+1}\}$.

RPFR reaches $t$ exactly when $b$ has a lower score than
every one of $x_1,\dots,x_{n-2}$. Each of these $n-1$ scores is equally likely
to be the smallest, so the probability is $1/(n-1)$.

Under target-first RPFR, the first reveal of $t$ happens when
$x_{n-3}$ is visited, and $t$ is then taken. Success holds exactly when $b$ has a
lower score than each of $x_1,\dots,x_{n-3}$; these labels plus $b$ number $n-2$,
so the probability is $1/(n-2)$.

Under \RDFS, survival at the source requires $x_1$ before $b$, with probability
$1/2$. At $x_1$ the visited source is skipped, and the relative order of $x_2$ and
$b$ is uniform and independent; since $n\ge5$, $x_2\ne t$, so survival requires
$x_2$ before $b$, again with probability $1/2$. After $x_2$ the DFS follows the
safe path and reaches $t$ before returning to a stack entry that can choose
$b$, with conditional probability $1$. The product is $1/4$.
\end{proof}

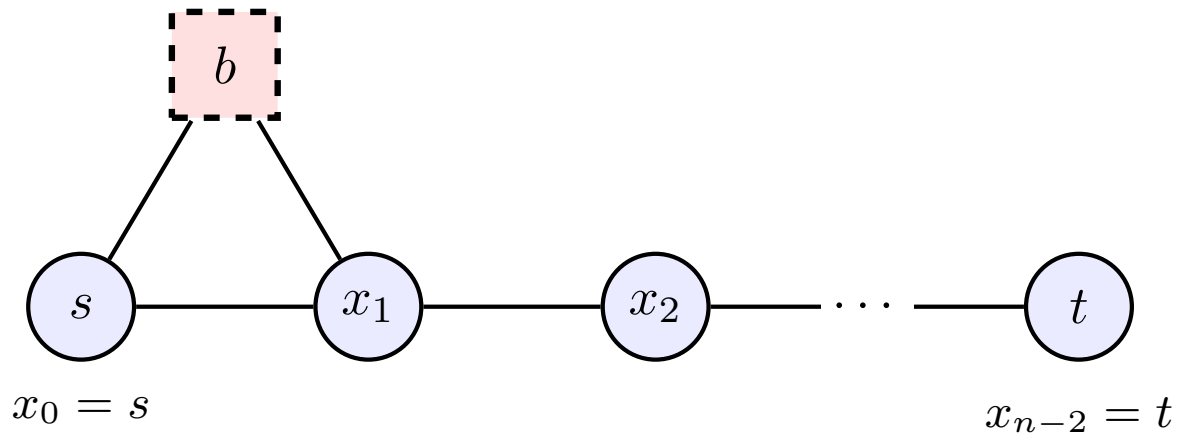
\begin{figure}[H]
\centering
\begin{tikzpicture}[>=Stealth, font=\small]
  \node[honest] (s) at (0,0) {$s$};
  \node[honest] (x1) at (1.9,0) {$x_1$};
  \node[honest] (x2) at (3.8,0) {$x_2$};
  \node[honest] (t) at (6.6,0) {$t$};
  \node[cartel] (b) at (0.95,1.6) {$b$};
  \node[tinylabel, below=1mm of s] {$x_0=s$};
  \node[tinylabel, below=1mm of t] {$x_{n-2}=t$};

  \draw[tree edge] (s) -- (x1);
  \draw[tree edge] (x1) -- (x2);
  \draw[tree edge] (x2) -- (t);
  \node[fill=white, inner sep=2pt] at (5.2,0) {$\cdots$};
  \draw[tree edge] (s) -- (b);
  \draw[tree edge] (x1) -- (b);

  \node[tinylabel, align=left, anchor=west] at (0.0,-1.5)
    {honest path $x_0,\dots,x_{n-2}$; cartel $b$ adjacent only to $s,x_1$};
  \node[tinylabel, align=left, anchor=west] at (0.0,-2.1)
    {maximum degree $3$; only visited rows are revealed};
\end{tikzpicture}

\smallskip
{\small
\begin{tabular}{@{}lccc@{}}
\toprule
$n$ & RPFR & target-first RPFR & \RDFS\\
 & $1/(n-1)$ & $1/(n-2)$ & $1/4$\\
\midrule
$5$ & $1/4$ & $1/3$ & $1/4$\\
$6$ & $1/5$ & $1/4$ & $1/4$\\
$7$ & $1/6$ & $1/5$ & $1/4$\\
\bottomrule
\end{tabular}}
\caption{The long-tail family: honest path $x_0=s,x_1,\dots,x_{n-2}=t$ plus one cartel vertex $b$ adjacent only to $s$ and $x_1$, with maximum degree $3$. The table gives exact success probabilities for $n=5,6,7$: RPFR decays as $1/(n-1)$, target-first RPFR as $1/(n-2)$, and \RDFS{} stays at $1/4$. Target-first RPFR is better at $n=5$, tied at $n=6$, and worse for $n\ge7$.}
\label{fig:priority-tail}
\end{figure}

Target-first RPFR is better than \RDFS{} at $n=5$, equal at $n=6$, and worse
at $n\ge7$; the $n$-dependent decay persists after the same override.
The family is the one-cartel specialisation of the
construction in Section~3 of \citet{petrucena2026random}; the target-first and
\RDFS{} values are proved here directly.

\subsection{A five-vertex worked example}\label{comparison:cactus}

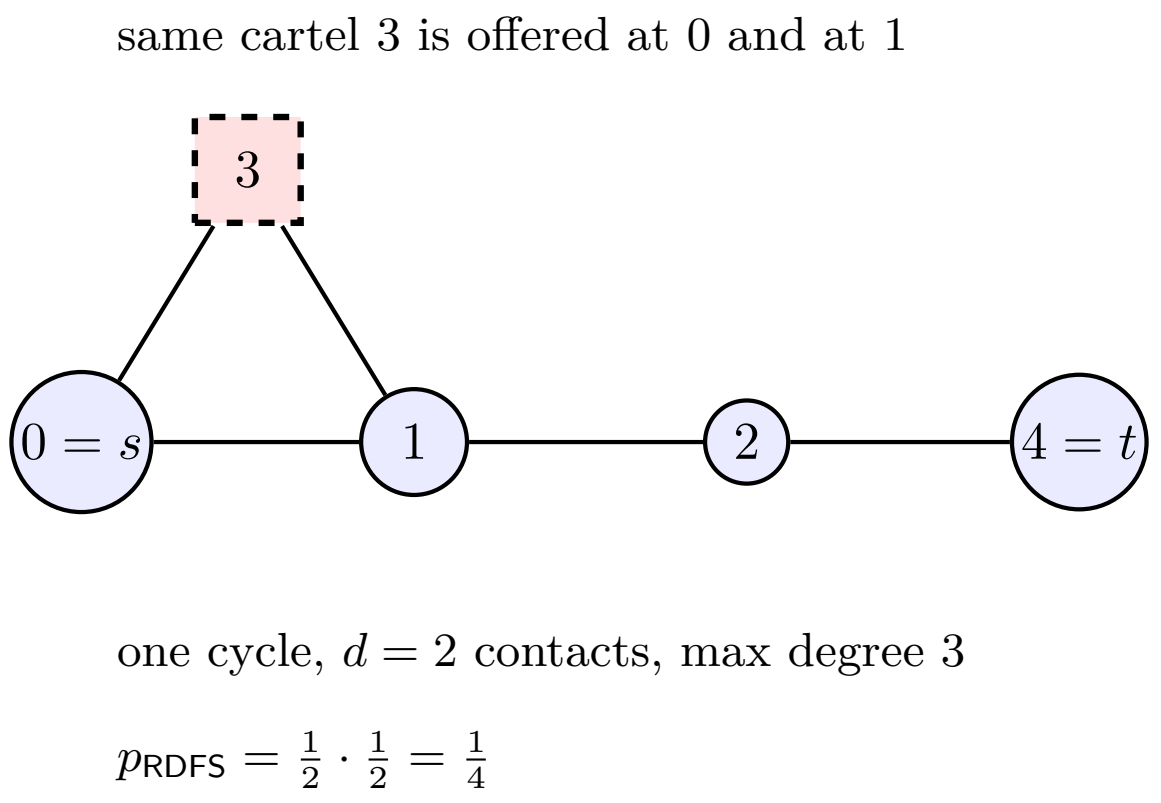
\begin{figure}[H]
\centering
\begin{tikzpicture}[>=Stealth, font=\small]
  \node[honest] (s) at (0,0) {$0=s$};
  \node[honest] (one) at (2.2,0) {$1$};
  \node[honest small] (buf) at (4.4,0) {$2$};
  \node[honest] (t) at (6.6,0) {$4=t$};
  \node[cartel] (b) at (1.1,1.8) {$3$};

  \draw[tree edge] (s) -- (one);
  \draw[tree edge] (one) -- (buf);
  \draw[tree edge] (buf) -- (t);
  \draw[tree edge] (s) -- (b);
  \draw[tree edge] (one) -- (b);

  \node[tinylabel, anchor=west] at (0.1,-1.4)
    {one cycle, $d=2$ contacts, max degree $3$};
  \node[tinylabel, anchor=west] at (0.1,-2.1)
    {$\pDFS=\frac12\cdot\frac12=\frac14$};
  \node[tinylabel, align=left, anchor=west] at (0.1,2.7)
    {same cartel $3$ is offered at $0$ and at $1$};
\end{tikzpicture}
\caption{The five-vertex example. \RDFS{} must choose $1$ before $3$ at the
source $0$, then $2$ before $3$ at vertex $1$, giving $\pDFS=1/4$. The same
compromised vertex $3$ is encountered as a candidate twice, so $d=2$.
Using $d=2$ in Theorem~\ref{thm:degree} gives $1/6$, below the actual value $1/4$.}
\label{fig:cactus-witness}
\end{figure}

The $n=5$ member has $E(G)=\{01,12,24,03,13\}$, with $s=0$, $t=4$ and
$C=\{3\}$. It is a triangle $0,1,3$ with the path $1,2,4$ attached, so $d=2$
and $\pDFS=1/4$ as in the family. Using $d=2$, the boundary form of the bound gives
$1/(3\cdot2)=1/6$: its root factor uses the cap $3$ rather than the actual
source degree $2$.
Figure~\ref{fig:cactus-witness} shows the instance.

\subsection{Ordinary random walks on a subdivided comb}\label{comparison:walks}

An ordinary random walk forgets where it has been, so every return to a risky
junction re-exposes a choice that \RDFS{} makes once and then leaves behind; the
family below makes that difference quantitative on a tree of maximum degree
three.

An \emph{ordinary random walk} (RW) starts at $s$, moves at each hop to a
uniformly chosen neighbour of the current vertex using fresh independent random
choices, may revisit vertices, and stops at the first vertex in $C\cup\{t\}$.

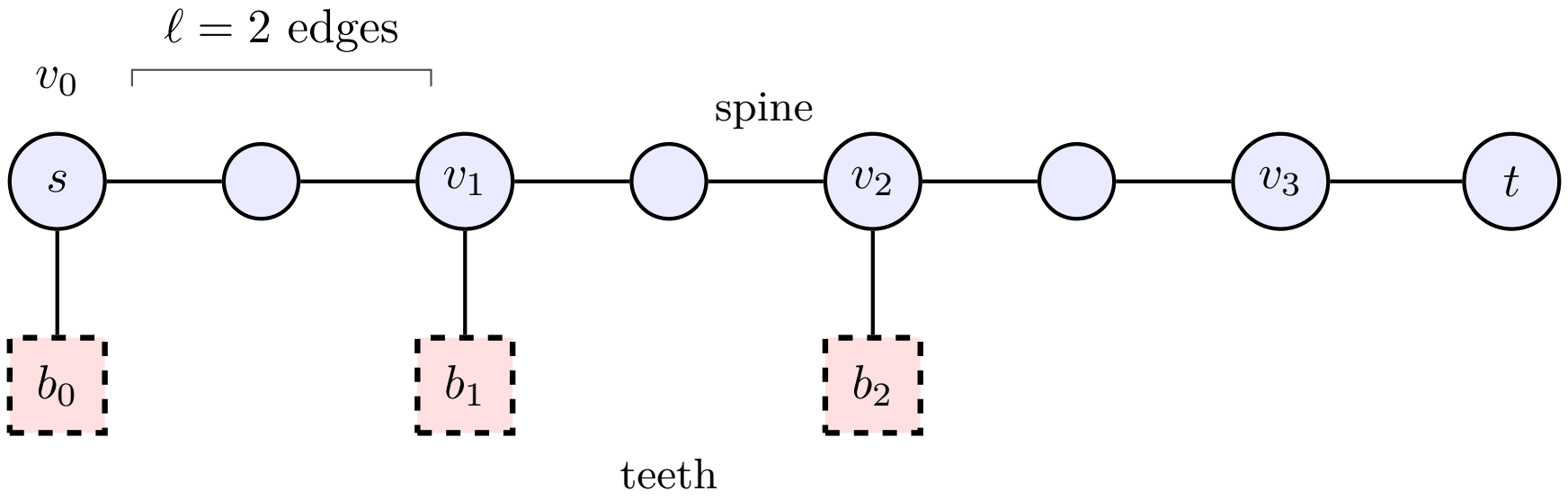
\begin{figure}[H]
\centering
\begin{tikzpicture}[>=Stealth, font=\small, x=1cm, y=1cm]
  \node[honest] (v0) at (0,0) {$s$};
  \node[honest] (v1) at (3,0) {$v_1$};
  \node[honest] (v2) at (6,0) {$v_2$};
  \node[honest] (v3) at (9,0) {$v_3$};
  \node[honest] (t)  at (10.7,0) {$t$};

  \node[honest small] (i1) at (1.5,0) {};
  \node[honest small] (i2) at (4.5,0) {};
  \node[honest small] (i3) at (7.5,0) {};

  \node[cartel] (b0) at (0,-1.5) {$b_0$};
  \node[cartel] (b1) at (3,-1.5) {$b_1$};
  \node[cartel] (b2) at (6,-1.5) {$b_2$};

  \draw[tree edge] (v0) -- (i1);
  \draw[tree edge] (i1) -- (v1);
  \draw[tree edge] (v1) -- (i2);
  \draw[tree edge] (i2) -- (v2);
  \draw[tree edge] (v2) -- (i3);
  \draw[tree edge] (i3) -- (v3);
  \draw[tree edge] (v3) -- (t);
  \draw[tree edge] (v0) -- (b0);
  \draw[tree edge] (v1) -- (b1);
  \draw[tree edge] (v2) -- (b2);

  \node[font=\small, above] at (0,0.5) {$v_0$};

  \draw[thin, black!70] (0.55,0.70) -- (0.55,0.82) --
        (2.75,0.82) -- (2.75,0.70);
  \node[font=\small, above] at (1.65,0.82) {$\ell=2$ edges};

  \node[tinylabel] at (5.2,0.52) {spine};
  \node[tinylabel] at (4.5,-2.15) {teeth};
\end{tikzpicture}
\caption{The subdivided comb $G_{3,2}$ ($n=11$, maximum degree $3$): a safe
spine with one compromised tooth at each of $v_0,v_1,v_2$. Each
junction-to-junction path has two edges. In the general family, lengthening
these paths creates more opportunities for a random walk to return to a risky
junction, but adds no new \RDFS{} branch choice.}
\label{fig:walk-comb}
\end{figure}

For integers $c\ge1$ and $\ell\ge1$, let $G_{c,\ell}$ have honest junctions
$v_0=s,v_1,\dots,v_c$; for $i<c$ an honest path $P_i$ of $\ell$ edges joins
$v_i$ to $v_{i+1}$, and the paths are pairwise internally disjoint. Add a target
$t$ adjacent only to $v_c$, and for $i<c$ a compromised leaf $b_i$ adjacent only
to $v_i$; put $C=\{b_0,\dots,b_{c-1}\}$. Then $G_{c,\ell}$ is a tree on
$n=c(\ell+1)+2$ vertices, its maximum degree is $3$ ($2$ when $c=1$), and
$G_{c,\ell}-C$ is the honest $s$-$t$ path. Figure~\ref{fig:walk-comb} draws
$G_{3,2}$.

\begin{samepage}
\begin{proposition}[A walk on a subdivided comb]\label{prop:walk-comb}
Let $a_0=1$, $a_1=\ell+1$ and $a_{j+1}=(\ell+2)a_j-a_{j-1}$ for $j\ge1$. For all
$c\ge1$ and $\ell\ge1$,
\begin{equation}\label{eq:walk-values}
\pDFS(G_{c,\ell},C)=2^{-c},\qquad
p_{\mathrm{RW}}(G_{c,\ell},C)=\frac{\ell}{(\ell+1)a_c-a_{c-1}}.
\end{equation}
For fixed $c$ as $\ell\to\infty$, $p_{\mathrm{RW}}$ is
$\Theta(\ell^{-c})=\Theta(n^{-c})$, with a constant depending on $c$, while
$\pDFS=2^{-c}$. At $\ell=1$, with $F_0=0$ and $F_1=1$, we have $a_j=F_{2j+1}$,
hence $p_{\mathrm{RW}}=1/F_{2c+2}$, and $p_{\mathrm{RW}}$ is
$\Theta(((3-\sqrt5)/2)^c)$ as $c\to\infty$, against $\pDFS=2^{-c}$.
\end{proposition}
\end{samepage}

\begin{proof}
The $c+1$ junctions, the $c(\ell-1)$ internal path vertices, the $c$ leaves and
$t$ give $n=c(\ell+1)+2$ vertices, and the graph is a tree. Its degrees are $2$
at $v_0$, $v_c$ and the internal path vertices, $3$ at $v_1,\dots,v_{c-1}$, and
$1$ at the leaves.

For \RDFS{}, each internal path vertex has only one unvisited neighbour, the
next vertex towards $t$. At each junction $v_i$ with $i<c$, the two unvisited
choices are the continuation and $b_i$. The search therefore either advances or
fails, without backtracking. Its random order at $v_i$ is independent of the
choices that brought it there, so the continuation comes first with probability
$1/2$. Multiplying over the $c$ junctions gives $\pDFS=2^{-c}$. Once $v_c$ is
reached, the target is visible and taken immediately.

For the ordinary walk, let $q(v)$ be the probability of reaching $t$ before $C$
when starting at $v$. Then $q(t)=1$ and $q(b_i)=0$. Conditioning on the first
step shows that at every other vertex, $q(v)$ is the average of its neighbours'
values.

Put $x_i=q(v_i)$ and write $w_{i,0}=v_i,\dots,w_{i,\ell}=v_{i+1}$ for the
vertices of $P_i$, endpoints included. Each internal $w_{i,k}$ has degree $2$, so
the averaging equation says that successive values differ by the same amount:
\[
q(w_{i,k})=x_i+\frac{k}{\ell}\,(x_{i+1}-x_i),\qquad 0\le k\le\ell ,
\]
where $k=0$ and $k=\ell$ restate the endpoint values; this also covers $\ell=1$,
when $P_i$ has no internal vertex. The neighbours of $v_0$ are $w_{0,1}$ and
$b_0$, so $2x_0=q(w_{0,1})$ and $x_1=(\ell+1)x_0$. The neighbours of $v_i$ for
$1\le i\le c-1$ are $w_{i-1,\ell-1}$, $w_{i,1}$ and $b_i$. For $\ell=1$ these
are $v_{i-1}$, $v_{i+1}$ and $b_i$; the averaging equation gives
\[
x_{i+1}=(\ell+2)x_i-x_{i-1}.
\]
The neighbours of $v_c$ are $w_{c-1,\ell-1}$ and $t$, so
\[
2x_c=\frac{x_{c-1}}{\ell}+\frac{\ell-1}{\ell}x_c+1,
\qquad\text{that is,}\qquad
(\ell+1)x_c=x_{c-1}+\ell .
\]
The first two relations give $x_j=a_jx_0$ for $0\le j\le c$, and the boundary
relation at $v_c$ gives
$p_{\mathrm{RW}}(G_{c,\ell},C)=x_0=\ell/((\ell+1)a_c-a_{c-1})$.

The sequence increases, because $a_0<a_1$ and
$a_{j+1}-a_j=(\ell+1)a_j-a_{j-1}>0$ whenever $a_j>a_{j-1}$. In particular, the
denominator above is positive. Since
$a_{j-1}\le a_j$, the recurrence also gives $a_{j+1}\ge(\ell+1)a_j$ and
$a_{j+1}<(\ell+2)a_j$, hence
\[
(\ell+1)^j\le a_j\le(\ell+2)^j \qquad (j\ge0).
\]
For $D=(\ell+1)a_c-a_{c-1}$, positivity and $a_{c-1}<a_c$ give
$\ell(\ell+1)^c\le\ell a_c<D<(\ell+1)(\ell+2)^c$, so
\[
\frac{\ell}{(\ell+1)(\ell+2)^c}
<p_{\mathrm{RW}}<\frac{1}{(\ell+1)^c}.
\]
With $n=c(\ell+1)+2$ and $c$ fixed, $\ell=\Theta(n)$, so the walk is
$\Theta(n^{-c})$, while $\pDFS=2^{-c}$.

At $\ell=1$ the recurrence is $a_{j+1}=3a_j-a_{j-1}$ with $a_0=1=F_1$ and
$a_1=2=F_3$; since $F_{2j+3}=3F_{2j+1}-F_{2j-1}$, induction gives
$a_j=F_{2j+1}$. Then $2F_{2c+1}-F_{2c-1}=F_{2c+2}$, so
$p_{\mathrm{RW}}=1/F_{2c+2}$. Binet's formula
$F_m=(\phi_+^m-\phi_-^m)/\sqrt5$, with $\phi_+=(1+\sqrt5)/2$ and
$\phi_-=(1-\sqrt5)/2$, gives $F_m=\Theta(\phi_+^m)$, and
$\phi_+^{-2}=(3-\sqrt5)/2<1/2$; the walk is therefore
$\Theta(((3-\sqrt5)/2)^c)$, while $\pDFS=2^{-c}$.

\end{proof}

The ordinary walk can return to a risky junction and face the same danger again;
\RDFS{} continues after each safe choice. Table~\ref{tab:walk-values} shows how
this difference grows with the lengths of the paths. On these examples, \RDFS{}
also attains the optimal tree guarantee $2^{-c}$ from Theorem~\ref{thm:trees},
since $n=c(\ell+1)+2\ge2c+2$.

\begin{table}[H]
\centering
\small
\begin{tabular}{@{}cccc@{}}
\toprule
$\ell$ & $n$ & $p_{\mathrm{RW}}$ & $\pDFS$\\
\midrule
$1$ & $10$ & $1/55$   & $1/16$\\
$2$ & $14$ & $1/209$  & $1/16$\\
$4$ & $22$ & $1/1189$ & $1/16$\\
$8$ & $38$ & $1/9701$ & $1/16$\\
\bottomrule
\end{tabular}
\caption{Exact success probabilities on $G_{4,\ell}$. Longer paths reduce the
ordinary walk's chance of survival, while \RDFS{} stays at $1/16$.}
\label{tab:walk-values}
\end{table}

\subsection{How many independent walkers?}\label{comparison:repetitions}

Suppose success means that at least one walker reaches $t$ safely. Let
$0<\varepsilon<1$ be the allowed probability that \emph{all} walkers fail.
Fix one admissible pair $(G,C)$ before any random choices, and run the same
policy $A$ independently $M$ times. Each walker starts at $s$ with fresh random
choices and its own routing memory; walkers do not share observations. Their
routes may overlap: independence concerns the random choices, not disjointness
of the routes.

If $p_A(G,C)\ge p_0>0$ is the one-run guarantee, then independence gives
\begin{equation}\label{eq:all-fail}
\Prob(\text{all $M$ walkers fail})
=(1-p_A(G,C))^M\le(1-p_0)^M.
\end{equation}
For $0<p_0<1$, solving $(1-p_0)^M\le\varepsilon$ shows that the least integer
count certified by this bound is
\begin{equation}\label{eq:walker-count}
M_\varepsilon(p_0)=
\left\lceil\frac{\log\varepsilon}{\log(1-p_0)}\right\rceil.
\end{equation}
Both logarithms are negative. If $p_0=1$, one walker suffices. This is the usual
independent-repetition calculation, also used for RPFR
\citep[Section~3.1]{petrucena2026random}. With
$p_0=W(\Delta,c)$, the count is independent of $n$ for fixed
$\Delta,c,\varepsilon$, although it grows rapidly with $c$.

Table~\ref{tab:walker-counts} gives sufficient counts from the stated one-run
guarantees. A particular graph may need fewer walkers if its actual survival
probability exceeds the guarantee. These counts are not lower bounds for
cooperating walkers or other multi-walker strategies.

\begin{table}[H]
\centering
\small
\renewcommand{\arraystretch}{1.15}
\begin{tabular}{@{}lccrrr@{}}
\toprule
Policy and graph class & $c$ & $p_0$ &
\multicolumn{3}{c}{$M_\varepsilon(p_0)$} \\
\cmidrule(l){4-6}
 & & & $\varepsilon=10^{-1}$ & $10^{-2}$ & $10^{-6}$ \\
\midrule
\RDFS{}, trees & $1$ & $1/2$ & $4$ & $7$ & $20$ \\
 & $2$ & $1/4$ & $9$ & $17$ & $49$ \\
 & $3$ & $1/8$ & $18$ & $35$ & $104$ \\
\addlinespace[0.5em]
\RDFS{}, degree cap $3$ & $1$ & $1/12$ & $27$ & $53$ & $159$ \\
 & $2$ & $1/96$ & $220$ & $440$ & $1320$ \\
 & $3$ & $1/768$ & $1768$ & $3535$ & $10604$ \\
\addlinespace[0.5em]
RPFR, $n=100$ & $1$ & $1/99$ & $227$ & $454$ & $1361$ \\
\bottomrule
\end{tabular}
\caption{Sufficient numbers of independent walkers for at least one safe arrival
with probability at least $1-\varepsilon$, respectively $90\%$, $99\%$ and
$99.9999\%$.
Tree rows use $p_0=2^{-c}$ for $n\ge2c+2$
(Theorem~\ref{thm:trees}); degree-three rows use $p_0=W(3,c)$ for every
admissible $n$ (Theorem~\ref{thm:degree}). The RPFR row uses its general
one-owl bound $1/(n-1)$. Every entry is the integer in
\eqref{eq:walker-count}, not a simulation estimate.}
\label{tab:walker-counts}
\end{table}

There is also an XOR-sharing interpretation. If these $M$ walkers carry $M$
different indexed shares, each sent once, Proposition~\ref{prop:xor-exposure}
shows that the same counts bound the probability of full share exposure by
$\varepsilon$, under its passive-observation and independent-routing
assumptions. This does not guarantee reconstruction, which needs all the
shares. Repeating the same share is different: one copy may expose it even if
another arrives safely.

\section{Related work}\label{sec:related-work}

The closest comparisons are relaying rules for the same hidden-topology task.
Other graph-search models differ in what is hidden, what failure means, or what
counts as success. We separate these comparisons from the proof techniques and
cryptographic motivation.

\subsection{The closest relaying rules}

This paper follows the random-priority frontier routing (RPFR) preprint
\citep{petrucena2026random}. RPFR fixes independent identically distributed
continuous priorities, equivalently a uniform random priority permutation, and
always visits the highest-priority frontier vertex without a target preference.
Its key guarantee, restated here for comparison, is that on every connected
simple graph with a fixed $c$-vertex cartel and an $s$-$t$ path avoiding the
cartel,
\[
p_{\mathrm{RPFR}}(G,C)\ge\binom{n-1}{c}^{-1}.
\]
The bound is attained by examples for that rule
\citep[Theorems~1 and~2]{petrucena2026random}. For fixed $c$ it decays like
$n^{-c}$; the quantity $\Theta(n^c)$ in the preprint's title counts independent
reruns, not the success probability of one run. Our degree-based guarantee is
instead independent of network size, and our proofs are self-contained. The
predecessor supplies a comparison strategy, not a premise of those proofs.

The highest-score (HS) heuristic in the published random-walk relaying study
\citep[Sections~3.2 and~3.8]{petrucena2026topology} gives each fragment a fresh
seed. It takes an adjacent target immediately; otherwise it compares fixed
seed-derived scores among unvisited neighbours of the current relay and uses a
non-backtracking fallback when none remain. It already retains a visited set; the
reported HS evidence is empirical. The RPFR preprint
\citep[Sections~2.1 and~4.2]{petrucena2026random} retains the
discovered frontier and enough topology to relocate through the visited region, so
a previously discovered, still-unvisited candidate remains eligible after a safe
detour. It compares priorities globally and gives the target no preference. The
preprint exhibits exponentially small success probability with one compromised
vertex for an idealised local HS rule with independent continuous scores
\citep[Proposition~4]{petrucena2026random}. That example has unbounded maximum
degree and does not establish the same claim for the hash-based HS rule of the
published study. The distinction is the retained frontier and the scope of each
choice.

\subsection{Other exploration objectives}

Incomplete information alone does not define one routing problem.
Table~\ref{tab:related-models} distinguishes the objectives of the closest
graph-search settings. In particular, discovering a blocked road does not
destroy the traveller, whereas entering a fatal vertex ends our walker's run.

\begin{table}[H]
\centering
\small
\renewcommand{\arraystretch}{1.15}
\begin{tabular}{@{}>{\raggedright\arraybackslash}p{0.25\textwidth}
                  >{\raggedright\arraybackslash}p{0.32\textwidth}
                  >{\raggedright\arraybackslash}p{\dimexpr0.43\textwidth-4\tabcolsep\relax}@{}}
\toprule
Problem & Information or danger & Objective \\
\midrule
Unknown-graph exploration \citep{deng1999exploring} &
Unknown strongly connected directed graph &
Traverse every edge, comparing travel cost with an explorer given the map. \\
\addlinespace[0.7em]
Canadian traveller \citep{papadimitriou1991shortest,barnoy1991canadian} &
Known road map; blockages discovered locally &
Reach a destination with low travel cost under uncertain road availability. \\
\addlinespace[0.7em]
Black-hole and dangerous-graph search \citep{dobrev2006searching,balamohan2016exploring} &
Harmful vertices can destroy visiting agents &
Identify danger or explore the graph using a cooperating team. \\
\addlinespace[0.7em]
This routing game &
Unknown graph; fixed fatal vertices; a safe route is promised &
Maximise the probability that one walker reaches the target before a fatal visit. \\
\bottomrule
\end{tabular}
\caption{Representative graph-search tasks differ in their information and success
criteria. Their travel-cost and team-size bounds do not directly determine our
single-walker survival value.}
\label{tab:related-models}
\end{table}

Bar-Noy and Schieber additionally consider bounded numbers of blockages,
recoverable roads and stochastic variants \citep{barnoy1991canadian}.
Black-hole search has the closer notion of danger: destruction on entry leaves
no observable trace. The asynchronous agents of Dobrev et al. communicate
through whiteboards and identify the black hole while at least one survives
\citep{dobrev2006searching}; Balamohan et al. study cooperation using movable
tokens \citep{balamohan2016exploring}. Our walker has neither a cooperating team
nor a requirement to identify the cartel or complete an exploration.

Childs et al. give an exponential quantum speedup for finding an initially
unknown exit label in a family of finite graphs of maximum degree three
\citep{childs2003exponential}. Their oracle can be queried on vertex labels
in superposition, and no vertex is fatal. This is a query-complexity result,
whereas our guarantee concerns a walker's survival while travelling to a known
target. Size-independent survival still permits up to $2(n-1)$ edge traversals
in one \RDFS{} run on an admissible instance.

\subsection{Depth-first order and distributional bounds}

Depth-first search is classical \citep{tarjan1972depth}. Gessel and Sagan
partition subgraphs by the deterministic DFS forest
\citep[Proposition~2.2]{gessel1995tutte}; here the randomness lies in the
neighbour order at each vertex, and Appendix~\ref{app:degree-proof} derives the
resulting conditional ordering law. Our upper bounds use the elementary
distributional argument underlying Yao's method \citep{yao1977probabilistic}:
bound every deterministic policy under a fixed distribution of instances, then
average over its internal randomness. The worst fixed instance is no easier
than that average. This direction needs no minimax equality.

\subsection{Secure transmission and planar routing}

The secret-sharing motivation connects to perfectly secure message transmission
\citep{dolev1993perfectly}, which seeks both privacy and reliable delivery despite
Byzantine faults under suitable network-connectivity assumptions. Trusted
repeater quantum key distribution networks similarly separate the privacy of
relayed key material from agreement and authentication at the endpoints
\citep[Sections~4.1-4.4]{salvail2010security}. These are protocol goals beyond
one packet's unexposed arrival. Our contribution at this interface is a routing
probability under unknown topology; Appendix~\ref{sec:bridge-shares} states
explicitly how independent routes combine with XOR sharing against passive
observation, and why reconstruction is a separate requirement.

Planarity is useful in some routing models, but placing a network in a physical
region does not make its connectivity graph planar. For example, GPSR constructs
a planar subgraph for its perimeter-routing mode
\citep[Section~2.3]{karp2000gpsr}. Here planarity has a different role: our lower
bound holds for all graphs under the degree cap, and planar matching examples
show that nonplanarity is not needed to attain the worst case.

\section{Conclusion and open problems}\label{sec:scope}

The main result is that local degree, rather than the total number of vertices,
can control worst-case survival in an unknown graph. Randomised depth-first
search achieves this by passing at most $d\le\Delta c$ risky branch points in
its analysis tree. The resulting guarantee is sharp for the rule and optimal
within strict DFS at every sufficiently large size, including on planar graphs.
The exact values on trees and on graphs of maximum degree two show what is
possible even for unrestricted policies. The comparisons with RPFR and ordinary
random walks also demonstrate that retaining information is not enough by
itself: the order in which that information is used matters.

For $\Delta\ge3$, determining the unrestricted values remains the main open
problem. For $n\ge2c+2$, the bounds proved here are
\[
W(\Delta,c)\le V(n,c,\Delta)\le V_{\mathrm{pl}}(n,c,\Delta)\le2^{-c}.
\]
At $\Delta=3$, the endpoints are $(2/3)8^{-c}$ and $2^{-c}$, leaving a gap in
the exponential dependence on $c$. Can a policy that switches between
unfinished branches improve on $W(\Delta,c)$ uniformly over all sufficiently
large $n$? The strict-DFS example does not settle this: its safe arm vertices
can be probed and left by such a policy. It is also open whether planarity
improves the unrestricted value, or gives a size-independent guarantee without
a degree cap.

The model fixes the graph and cartel before the random choices, gives truthful
neighbour information on safe visits, and caps the degrees of cartel vertices
as well as safe ones. The theorem uses exact independent local orders;
Section~\ref{model:seeded} describes only a computational implementation proposal.
If $c$ grows with $n$, the survival guarantee may decrease. A single \RDFS{} run
can be implemented with at most $2(n-1)$ edge traversals, so size-independent
survival does not mean size-independent communication cost.

For secret sharing, independent routing gives the passive XOR-exposure bound in
Proposition~\ref{prop:xor-exposure}. A useful end-to-end protocol must additionally
ensure that enough authentic shares arrive, and must account for any dependence
introduced by retries or shared routing state. Quantifying that delivery and
privacy tradeoff is distinct from maximising one walker's survival probability.

Finally, \RDFS{} remembers visited vertices, a search stack and local neighbour
orders. What survival probability can the hedgehog guarantee if its memory is
limited? In particular, for fixed $\Delta$ and $c$, how much memory is needed
to keep the guarantee independent of $n$?

\appendix
\section{Proof of the size-independent bound}\label{app:degree-proof}

This appendix proves Theorem~\ref{thm:degree} and completes the verification of
Theorem~\ref{thm:sharp}. The graph $H$, the marked leaves $\mathcal{L}$, and the boundary
size $d$ are those of Section~\ref{sec:model}. All permutations are independent
and uniform on the whole neighbour set of the vertex where they are drawn. A
\emph{port} of $v$ is one entry in its neighbour order $\pi_v$; scanning the order
means inspecting these entries one by one. A vertex is \emph{visited} only when
the search has entered it; being named in a revealed row does not make it visited.

\subsection{Conventions, termination and tree structure}\label{app:dfs}

We first show that the search ends. The key is that it skips vertices already
visited, rather than repeatedly exploring them.

\begin{lemma}[Termination]\label{lem:termination}
On every finite instance the rule \RDFS{} stops after finitely many steps. On an
admissible instance it stops at the target or at a cartel vertex before the stack
can empty.
\end{lemma}

\begin{proof}
Each push enters a vertex not visited before, because visited labels are skipped;
there are at most $n$ pushes, at most $n$ pops, and each row has finite length, so
the process stops. Suppose the stack empties without a target visit or a cartel
entry, and let $S$ be the visited set. Every $v\in S$ lies in $K$, because the
walker reached $v$ through safe vertices. If $S\ne K$, a shortest path in $K$
from $S$ to $K\setminus S$ ends in an edge $vu$ with $v\in S$ and
$u\in K\setminus S$. The row of $v$ was exhausted, so $u$ was scanned and, being
unvisited, would have been taken. Hence $S=K$. But $t\in K$, and the rule stops on
entering $t$, a contradiction. Thus on an admissible instance the fallback is not
reached before success or failure.
\end{proof}

Now run plain DFS on a connected simple undirected graph $H$, treating every
vertex as safe. Let $T$ record the edge by which each vertex was first entered.
This tree is rooted at $s$ and spans $H$, but does not record the order in which
children were explored. Write $T_u$ for the subtree rooted at $u$, including $u$.

\begin{lemma}[No cross edges]\label{lem:nocross}
For every edge $xy$ of $H$, one endpoint is an ancestor of the other in $T$.
\end{lemma}

\begin{proof}
In a recursive DFS each vertex has a call interval, open at first visit and closed
when popped after its reachable unvisited region is exhausted. Intervals of
incomparable vertices are disjoint; ancestor and descendant intervals are nested.
Suppose an edge $xy$ joins incomparable $x,y$, with $x$ visited first. The frame
of $x$ closes before $y$ is first visited. When $x$ scanned $xy$, $y$ was still
unvisited; otherwise $y$ would have been visited earlier. Hence DFS recursed to
$y$ inside the frame of $x$, making $y$ a descendant of $x$, a contradiction.
\end{proof}

Thus every neighbour of $v$ is either an ancestor of $v$ or lies in $V(T_u)$ for
exactly one child $u$ of $v$. For each child $u$ of $v$ define the \emph{port
group}
\begin{equation}\label{app:ports}
A_{v,u}=N_H(v)\cap V(T_u),\qquad w_{v,u}=|A_{v,u}|\ge1,
\end{equation}
and let $A_{v,0}$ be the ancestor neighbours of $v$. These groups collect
neighbours of $v$; their positions in the local order $\pi_v$ are the ports of
$v$. By Lemma~\ref{lem:nocross} the sets $A_{v,u}$ over all children $u$, together
with $A_{v,0}$, partition $N_H(v)$. The weight $w_{v,u}$ counts neighbours of $v$
in $T_u$, not the size of $T_u$. Ancestor ports are already visited and are
ignored.

\begin{lemma}[Tree characterisation]\label{lem:tree-char}
For every rooted spanning tree $T$ with $\Prob(T)>0$, the event
$\{\text{DFS tree}=T\}$ is exactly
\begin{equation}\label{app:eq-ET}
\text{for every vertex $v$ and every child $u$ of $v$: }
u \text{ is the first element of } A_{v,u} \text{ in } \pi_v.
\end{equation}
\end{lemma}

\begin{proof}
Fix such a $T$. Since $\Prob(T)>0$, choose an outcome of the random choices
realising $\{\text{DFS tree}=T\}$ and apply Lemma~\ref{lem:nocross} to that
outcome. Thus this fixed $T$ has no $H$-edge with incomparable endpoints, a
structural property of $T$ independent of the current permutations.

Suppose the DFS tree is $T$. Fix $v$ and a child $u$ of $v$.
Before $u$ is first visited, every vertex of $T_u$ is unvisited: if $y\in T_u$
had been visited earlier, its first-visit parent chain would pass through $u$, so
$u$ would be earlier. If some $x\in A_{v,u}$, $x\ne u$, occurred before $u$ in
$\pi_v$, then when $v$ scanned $x$, no vertex of $T_u$ had been visited, so $x$
was unvisited and $v$ would recurse into $x$, making $x$ a child of $v$. This
contradicts $x$ lying in the subtree of $u$. Hence $u$ is first in $A_{v,u}$.

For the other direction, assume \eqref{app:eq-ET} and the no-cross property of this
fixed $T$, and prove by structural induction on $T$ that the DFS tree is $T$.
Maintain the invariant that when the DFS is about to process a child group of
$v$: all ancestors of $v$ are visited; every child subtree of $v$ not yet entered
is entirely unvisited; and every edge from a child subtree of $v$ to its
complement ends at $v$ or at an ancestor of $v$. The boundary clause is the
no-cross property of the fixed $T$, not a consequence of the visited set;
initially at the root only $s$ is visited.

Let $u$ be a child of $v$. By \eqref{app:eq-ET}, the first port of $v$ in
$A_{v,u}$ is $u$, and $u$ is unvisited then. The DFS may first process other
child groups of $v$; by the induction hypothesis, exploring such a child exhausts
only that child's subtree, so the first designated port $u$ remains unvisited
until its group is entered. We claim that entering $u$ explores exactly $T_u$. An
edge from $x\in T_u$ to outside $T_u$ ends at $v$ or at an ancestor of $v$ by the
boundary clause, and is already visited and skipped. There is no edge to a
sibling subtree, whose endpoints would be incomparable. Inside $T_u$,
\eqref{app:eq-ET} holds by inheritance, and the same induction applies with the
parent of $u$ as an already visited ancestor. Thus the DFS explores all of $T_u$
and returns to $v$ only after $T_u$ is exhausted.

After all child groups of $v$ are processed, every port of $v$ lies in an
exhausted child subtree or is an ancestor, hence visited. The DFS backtracks from
$v$, and the parent edges produced are exactly those of $T$. Since $H$ is
connected and $T$ is spanning, the DFS tree is $T$.
\end{proof}

\subsection{How neighbour groups are ordered}\label{app:port}

Give each group $A_{v,u}$, including $A_{v,0}$, a different colour. A uniform
neighbour order can be generated in two independent stages: randomly order the
labels within each colour, then uniformly shuffle all the colours together.
There are $w_{v,u}$ copies of child colour $u$ and $|A_{v,0}|$ copies of the
ancestor colour. We call this sequence of colours the \emph{interleaving word}.

\begin{lemma}[Conditional factorisation]\label{lem:cond-factor}
Fix $T$ with $\Prob(T)>0$ and condition on $\{\text{DFS tree}=T\}$. For every
vertex $v$, the interleaving word at $v$ is still uniform and is independent of
the internal orders at $v$. The interleaving words at different vertices are
independent.
\end{lemma}

\begin{proof}
The event \eqref{app:eq-ET}, restricted to $v$, says only that $u$ is first in
its own group $A_{v,u}$. It constrains internal orders alone, not the colour
interleaving. Under this condition each internal order is uniform over orders
with the prescribed first element, while the interleaving word remains uniform
and independent. Different vertices use independent $\pi_v$, and each constraint
at $v$ involves only $\pi_v$, so interleaving words are independent across
vertices.
\end{proof}

\begin{corollary}[First-group factor]\label{cor:first-factor}
For a distinguished child $g$ of $v$ and a set $B$ of other children, the
relative order of $\{A_{v,g}\}\cup\{A_{v,b}:b\in B\}$ in a uniform interleaving
is uniform over that sub-multiset. Hence the first element is uniform over its
$w_{v,g}+\sum_{b\in B}w_{v,b}$ individual positions, and
\begin{equation}\label{app:first}
\Prob\bigl(g\text{ is first among }\{g\}\cup B \bigm| T\bigr)
=\frac{w_{v,g}}{w_{v,g}+\sum_{b\in B}w_{v,b}}.
\end{equation}
Harmless groups and ancestor ports do not affect the relative order inside the
union. This is the only child-order probability used below. The induced order is
\emph{not} uniform: two children of weights $1$ and $2$ are first with
probabilities $1/3$ and $2/3$, not $1/2$ each. Figure~\ref{fig:port-groups}
below illustrates the law.
\end{corollary}

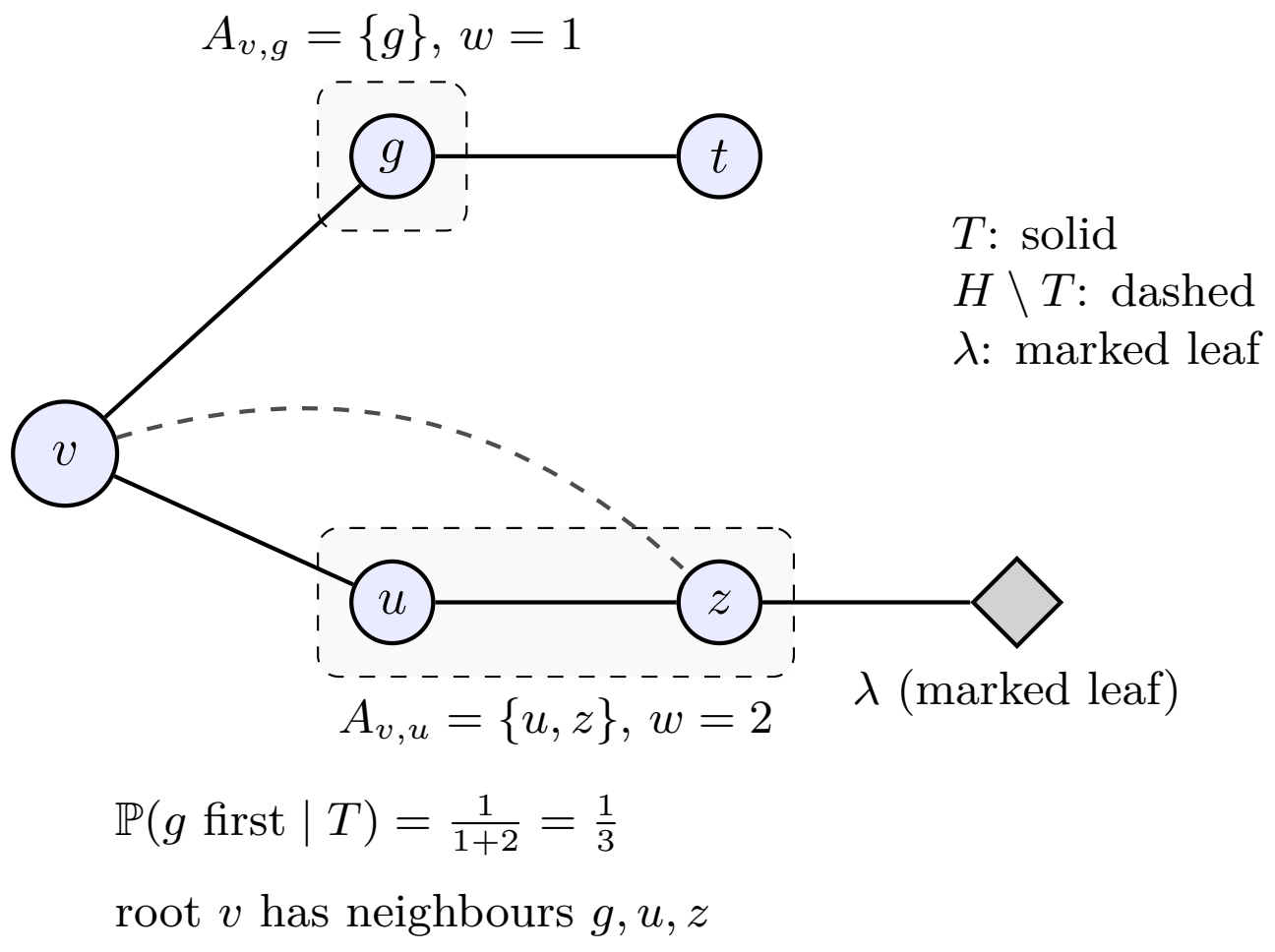
\begin{figure}[H]
\centering
\begin{tikzpicture}[>=Stealth, font=\small]
  \node[honest] (v) at (0,0) {$v$};
  \node[honest small] (g) at (2.2,2.0) {$g$};
  \node[honest small] (t) at (4.4,2.0) {$t$};
  \node[honest small] (u) at (2.2,-1.0) {$u$};
  \node[honest small] (z) at (4.4,-1.0) {$z$};
  \node[marked leaf, label={[tinylabel]below:$\lambda$ (marked leaf)}] (lam)
        at (6.4,-1.0) {};

  \draw[tree edge] (v) -- (g);
  \draw[tree edge] (g) -- (t);
  \draw[tree edge] (v) -- (u);
  \draw[tree edge] (u) -- (z);
  \draw[tree edge] (z) -- (lam);
  \draw[back edge] (v) to[bend left=30] (z);

  \begin{scope}[on background layer]
  \node[draw, dashed, rounded corners, fill=gray!5, fit=(g), inner sep=6pt,
        label={[tinylabel]above:$A_{v,g}=\{g\}$, $w=1$}] {};
  \node[draw, dashed, rounded corners, fill=gray!5, fit=(u)(z), inner sep=6pt,
        label={[tinylabel]below:$A_{v,u}=\{u,z\}$, $w=2$}] {};
  \end{scope}

  \node[tinylabel, align=left] at (7.0,1.1)
    {$T$: solid\\ $H\setminus T$: dashed\\ $\lambda$: marked leaf};
  \node[tinylabel, align=left, anchor=west] at (0.2,-3.1)
    {root $v$ has neighbours $g,u,z$};
  \node[tinylabel, align=left, anchor=west] at (0.2,-2.5)
    {$\Prob(g\text{ first}\mid T)=\frac{1}{1+2}=\frac13$};
\end{tikzpicture}
\caption{Two child subtrees, but three neighbouring vertices. Conditional on the
shown DFS tree $T$, the safe group $\{g\}$ has weight $1$ and the risky group
$\{u,z\}$ has weight $2$. Thus $\Prob(g\text{ first}\mid T)=1/3$, not $1/2$:
counting children equally would give the wrong answer.}
\label{fig:port-groups}
\end{figure}

\subsection{Counting risky branch points}\label{app:leaf}

Let $H$ be connected and simple, with source $s$, honest target $t\ne s$, and a
set $\mathcal{L}$ of $d\ge1$ marked leaves, $s,t\notin\mathcal{L}$. A
\emph{leaf} has degree $1$; the target may be an honest leaf but is not marked.
Run the plain all-safe DFS rooted at $s$, and let $T$ be its tree.

\begin{lemma}[Marked leaves are tree leaves]\label{lem:leaf-fact}
Every marked leaf is a leaf of $T$. Hence the $T$-path from $s$ to $t$ contains
no vertex of $\mathcal{L}$.
\end{lemma}

\begin{proof}
For a marked leaf $x\in\mathcal{L}$, $x\ne s$ because
$s\notin\mathcal{L}$, so its unique neighbour in $H$ is necessarily its parent
in $T$; a child would require another neighbour. Hence $x$ is a leaf of $T$. A
leaf of $T$ cannot lie on the $s$-$t$ path unless it is the endpoint $t$, since
every other vertex of that path has a child on the path, and $t$ is not marked.
\end{proof}

Let $\operatorname{Anc}_T(t)$ be the set of \emph{proper} ancestors of $t$ in
$T$. Success is decided at the first visit to $t$, before the row of $t$ is read,
so the row of $t$ and any marked leaves below $t$ are irrelevant; this is why $t$
is excluded from $\operatorname{Anc}_T(t)$. For
$v\in\operatorname{Anc}_T(t)$, let $g(v)$ be the child of $v$ on the $T$-path to
$t$, the \emph{good child}, and put
\begin{equation}\label{app:Bv}
B_v=\bigl\{\,u:\ u\text{ is a child of }v,\ u\ne g(v),\
V(T_u)\cap\mathcal{L}\ne\emptyset\,\bigr\}.
\end{equation}
Call $v$ \emph{risky} if $B_v$ is nonempty, and let $R$ be the set of risky
vertices.

\begin{lemma}[Leaf budget]\label{lem:leaf-budget}
$\sum_{v\in\operatorname{Anc}_T(t)}|B_v|\le d$, and consequently $|R|\le d$.
\end{lemma}

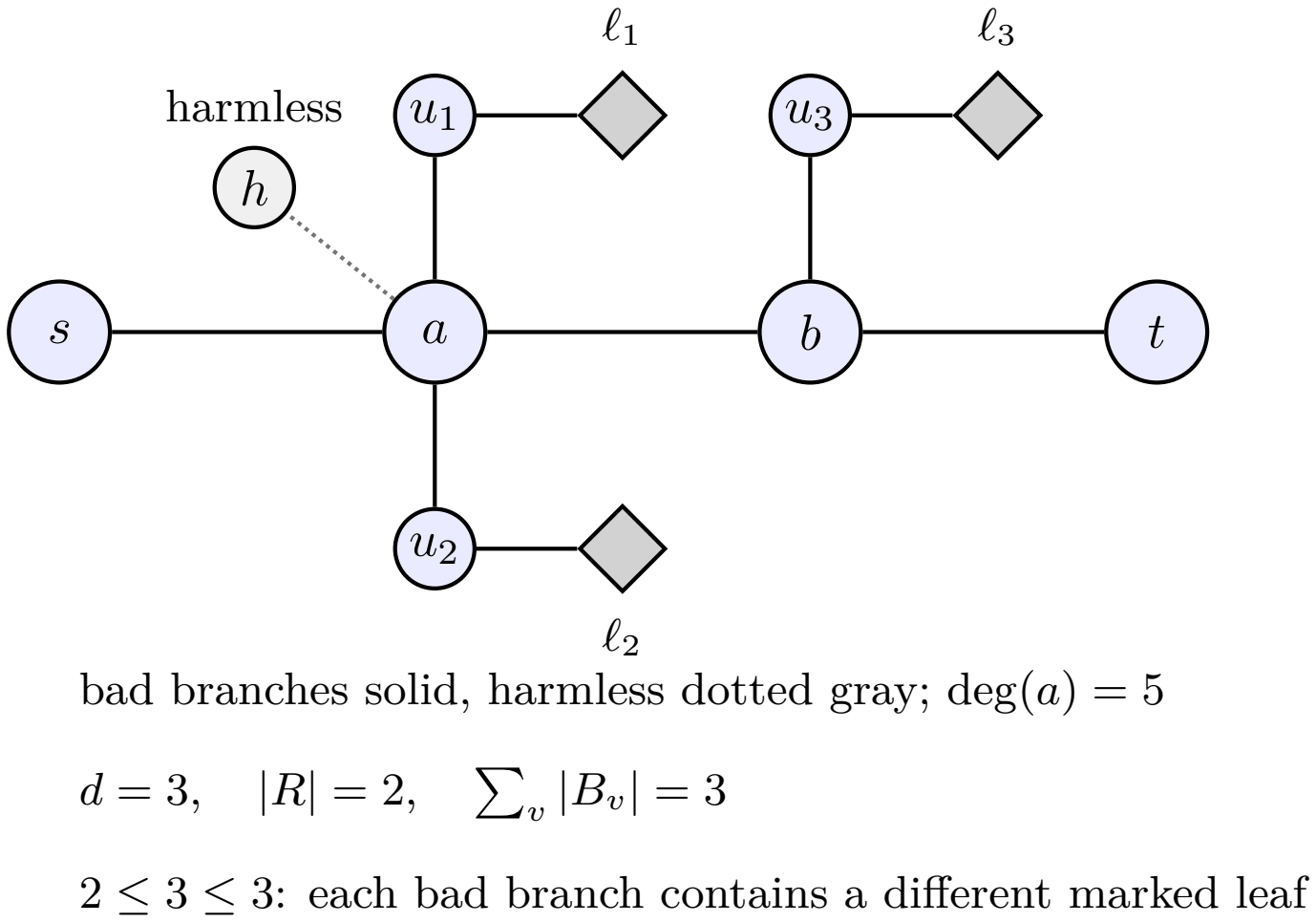
\begin{figure}[H]
\centering
\begin{tikzpicture}[>=Stealth, font=\small,
    badbranch/.style={thick, black},
    harmlessbranch/.style={thick, black!55, densely dotted}]
  \node[honest] (s) at (0,0) {$s$};
  \node[honest] (a) at (2.6,0) {$a$};
  \node[honest] (b) at (5.2,0) {$b$};
  \node[honest] (t) at (7.6,0) {$t$};
  \draw[tree edge] (s) -- (a);
  \draw[tree edge] (a) -- (b);
  \draw[tree edge] (b) -- (t);

  \node[honest small] (u1) at (2.6,1.5) {$u_1$};
  \node[marked leaf] (l1) at (3.9,1.5) {};
  \node[honest small] (u2) at (2.6,-1.5) {$u_2$};
  \node[marked leaf] (l2) at (3.9,-1.5) {};
  \draw[badbranch] (a) -- (u1); \draw[badbranch] (u1) -- (l1);
  \draw[badbranch] (a) -- (u2); \draw[badbranch] (u2) -- (l2);
  \node[tinylabel, above=1pt of l1] {$\ell_1$};
  \node[tinylabel, below=1pt of l2] {$\ell_2$};

  \node[honest small, fill=gray!12] (h) at (1.35,1.0) {$h$};
  \draw[harmlessbranch] (a) -- (h);
  \node[tinylabel, above=1pt of h] {harmless};

  \node[honest small] (u3) at (5.2,1.5) {$u_3$};
  \node[marked leaf] (l3) at (6.5,1.5) {};
  \draw[badbranch] (b) -- (u3); \draw[badbranch] (u3) -- (l3);
  \node[tinylabel, above=1pt of l3] {$\ell_3$};

  \node[tinylabel, align=left, anchor=west] at (0.0,-2.5)
    {bad branches solid, harmless dotted gray; $\deg(a)=5$};
  \node[tinylabel, align=left, anchor=west] at (0.0,-3.2)
    {$d=3,\quad |R|=2,\quad \sum_{v}|B_v|=3$};
  \node[tinylabel, align=left, anchor=west] at (0.0,-3.9)
    {$2\le3\le3$: each bad branch contains a different marked leaf};
\end{tikzpicture}
\caption{Counting risky branch points in the plain-DFS tree. The three disjoint
bad branches contain different marked leaves, while the harmless branch does
not affect survival. Thus $|R|=2\le\sum_v|B_v|=3\le d=3$. This count concerns
plain DFS, before adding the instruction to take a revealed target immediately.}
\label{fig:leaf-charging}
\end{figure}

\begin{proof}
The bad subtrees $T_u$, for all $v\in\operatorname{Anc}_T(t)$ and
$u\in B_v$, are pairwise vertex-disjoint. For distinct
$v,w\in\operatorname{Anc}_T(t)$, both lie on the $s$-$t$ path, so one is an
ancestor of the other. If $v$ is a proper ancestor of $w$, then $w$ lies in
$T_{g(v)}$ or $w=g(v)$. Every bad subtree at $w$ is contained in $T_{g(v)}$,
while every bad subtree at $v$ is disjoint from $T_{g(v)}$; iterating gives
disjointness. Each bad subtree contains a marked leaf, and disjoint subtrees
give distinct marked leaves. With $d$ marked leaves,
$\sum_{v\in\operatorname{Anc}_T(t)}|B_v|\le d$. Since every risky vertex
contributes at least one bad child, $|R|\le d$.
Figure~\ref{fig:leaf-charging} shows the accounting.
\end{proof}

Let $\mathrm{Success}$ be the event that the plain DFS visits $t$ before any
vertex of $\mathcal{L}$.

\begin{lemma}[Conditional leaf law]\label{lem:leaf-law}
For every $T$ with $\Prob(T)>0$,
\begin{equation}\label{app:leafeq}
\Prob(\mathrm{Success}\mid T)
=\prod_{v\in\operatorname{Anc}_T(t)}
\frac{w_{v,g(v)}}{w_{v,g(v)}+\sum_{b\in B_v}w_{v,b}},
\end{equation}
with the factor interpreted as $1$ when $B_v$ is empty.
\end{lemma}

\begin{proof}
By Lemma~\ref{lem:cond-factor}, conditional on $T$ the interleaving words are
independent and uniform; internal orders do not enter the success event beyond
first-in-group.

Let $G_v=\{g(v)\text{ is first among }\{g(v)\}\cup B_v\}$ at $v$. We claim
$\mathrm{Success}$ is exactly $\bigwedge_{v\in\operatorname{Anc}_T(t)}G_v$.
If some $G_v$ fails, a bad child $u\in B_v$ occurs before
$g(v)$. The DFS enters $u$ before $g(v)$, explores $T_u$, which contains a marked
leaf, before returning. Since $u\ne g(v)$, $t\notin T_u$, so a marked leaf
precedes $t$. Conversely, assume all $G_v$ and let the $T$-path be
$s=v_0,\dots,v_m=t$. Induct on $i$: $v_i$ is visited before any $\mathcal{L}$-vertex.
Suppose this holds for $v_i$; by $G_{v_i}$ the first child among $g(v_i)=v_{i+1}$
and all bad children is $v_{i+1}$. Any harmless child processed first contains
no marked leaf and no $t$, and returns without visiting either. Then $v_{i+1}$ is
entered before any bad child, still with no $\mathcal{L}$-vertex. At $i=m$, $t$ is visited
before any marked leaf.

The events $G_v$ depend on interleaving words at distinct vertices, so they are
conditionally independent, and $\Prob(G_v\mid T)$ is the ratio in
\eqref{app:first}. Multiplying gives \eqref{app:leafeq}.
\end{proof}

\begin{lemma}[Leaf bound]\label{lem:leaf-bound}
For every $T$ with $\Prob(T)>0$,
\begin{equation}\label{app:bound}
\Prob(\mathrm{Success}\mid T)
\ge
\frac{1}{\Delta(\Delta-1)^{d-1}},
\end{equation}
where $\Delta\ge2$ is an upper bound on the maximum degree of $H$ and $d=|\mathcal{L}|$.
\end{lemma}

\begin{proof}
Non-risky factors are $1$. For risky $v$, the numerator is $w_{v,g(v)}\ge1$ and
the denominator is at most the total child-group weight of $v$, which is
$\deg_H(v)$ minus the number of ancestor neighbours. For $v\ne s$ the parent is an
ancestor neighbour, so this is at most $\Delta-1$ and the factor is at least
$1/(\Delta-1)$. For $v=s$ it is at most $\Delta$, giving $1/\Delta$. There are at
most $d$ risky vertices.

If the root is risky, the product is at least
$\frac{1}{\Delta}(\frac{1}{\Delta-1})^{|R|-1}\ge
\frac{1}{\Delta}(\frac{1}{\Delta-1})^{d-1}$, since $|R|-1\le d-1$ and
$1/(\Delta-1)\le1$. If the root is not risky, it is at least
\[
\Bigl(\frac{1}{\Delta-1}\Bigr)^{|R|}
\ge
\Bigl(\frac{1}{\Delta-1}\Bigr)^{d}
\ge
\frac{\Delta-1}{\Delta}
\Bigl(\frac{1}{\Delta-1}\Bigr)^{d}
=
\frac{1}{\Delta(\Delta-1)^{d-1}},
\]
using $|R|\le d$ and $(\Delta-1)/\Delta\le1$. The case $|R|=0$ is included, since
$1$ exceeds the bound. For $\Delta=2$ both cases give at least $1/2$.
\end{proof}

\begin{corollary}[Averaged leaf bound]\label{cor:leaf-average}
Averaging \eqref{app:bound} over $T$,
\begin{equation}\label{app:avg}
\Prob(\mathrm{Success})\ge\frac{1}{\Delta(\Delta-1)^{d-1}}.
\end{equation}
\end{corollary}

\subsection{Replacing fatal contacts by leaves}\label{app:clone}

The next step turns fatal contacts into an all-safe exploration problem: each
contact becomes a marked leaf, so a fatal step is replaced by reaching a
distinguished vertex, while every safe degree and local order stays the same. Fix
an admissible pair $(G,C)$ and let $K$ be the source component of $G-C$. It
contains $t$, and every vertex visited before a fatal entry into $C$ lies in $K$. Every neighbour of $K$ outside $K$ lies in
$C$; otherwise two components of $G-C$ would be connected. Since $G$ is connected
and $C\ne\emptyset$, $d=|E(C,K)|\ge1$.

The analysis graph $H$ is built from $G[K]$. For each incidence $(v,b)$ with
$v\in K$, $b\in C$ and $vb\in E(G)$, add a \emph{new} vertex $\ell_{v,b}$, a
marked leaf, and the edge $v\ell_{v,b}$. Discard all other components of $G-C$
and all interior structure of $C$. Set
\begin{equation}\label{app:L}
\mathcal{L}=\{\ell_{v,b}:vb\in E(C,K)\},\qquad |\mathcal{L}|=d.
\end{equation}
Distinct incidences get distinct fresh leaves even when they share $b$.

\begin{lemma}[Properties of $H$]\label{lem:h-facts}
The graph $H$ is connected and simple. For every $v\in K$,
$\deg_H(v)=\deg_G(v)\le\Delta$, and every leaf in $\mathcal{L}$ has degree
$1$. The set $\mathcal{L}$ consists of $d$ marked leaves, and
$s,t\notin\mathcal{L}$.
\end{lemma}

\begin{proof}
$K$ is connected, and attaching one new leaf to each boundary incidence preserves
connectedness and simplicity. The degree identity holds because every neighbour
of $v$ in $G$ is either in $K$ or in $C$, and each $C$-neighbour is replaced by
exactly one new leaf. Leaves are new and have degree $1$. The source and target
lie in $K$ and are not new leaves.
\end{proof}

Let $p_{\mathrm{plain}}(G,C)$ be the probability that the plain DFS on $(G,C)$
visits $t$ before any vertex of $C$. Let $\pi_v$ be the local permutation used by
that DFS at a safe vertex $v\in K$. Define a bijection
\begin{equation}\label{app:phi}
\phi_v:N_G(v)\to N_H(v)
\end{equation}
that is the identity on $N_K(v)$ and maps $b\in N_C(v)$ to $\ell_{v,b}$. Set
$\pi^H_v=\phi_v(\pi_v)$. Since $\pi_v$ is uniform, so is $\pi^H_v$, and different $v$
remain independent; drawing an order in advance has the same effect as drawing it
on first arrival. Leaf
permutations in $H$ are degree-$1$ trivial, and the full all-safe $H$ tree may use
independent random choices after the stopping time, which is analysis only. The
rule uses only the terminal roles, not the values of $n$ or $c$, so relabelling
$H$ with source $0$ and target last does not change the uniform port law.

\begin{lemma}[Stopped coupling]\label{lem:coupling}
Run the plain DFS on $G$ and stop it at the first visit to $C$ (failure) or to
$t$ (success). Run the plain all-safe DFS on $H$ with the coupled permutations
and stop it at the first visit to $\mathcal{L}$ (failure for the comparison) or to $t$
(success). The two stopped runs have the same sequence of visited $K$-vertices,
the same sequence of scanned ports up to the bijection $\phi_v$, and the same
success or failure outcome. Consequently
\begin{equation}\label{app:couple}
p_{\mathrm{plain}}(G,C)
=
\Prob_H(t\text{ is visited before }\mathcal{L}).
\end{equation}
\end{lemma}

\begin{proof}
Induct on steps. The DFS stack and the visited subset of $K$ agree. Before the
stop, no vertex of $C$ and no vertex of $\mathcal{L}$ has been visited, whatever labels
earlier safe rows revealed, so no forbidden port is skipped as visited under the
replacement. At $v\in K$ the next unscanned
entry of $\pi_v$ maps to the next unscanned entry of $\pi^H_v$. If it is in
$K\setminus\{t\}$, it is the same vertex, with the same visited status. If it is
$b\in C$, then $G$ stops fatally while $H$ stops at the leaf $\ell_{v,b}\in\mathcal{L}$,
the same failure outcome. If it is $t$, both stop successfully; $t\in K$ and
$t\notin C\cup\mathcal{L}$, though it may be an honest degree-$1$ vertex. There is no
other case, since every neighbour of $K$ outside $K$ is in $C$.

Agreement therefore holds until the common stopping time. After a leaf is entered
in $H$, the all-safe DFS would return, but the stopped run stops at the
first-$\mathcal{L}$ time; a full all-safe $H$-run visits its first
$\mathcal{L}$ at the same time, because a leaf has no descendants. Hence the
success event ``$t$ before $\mathcal{L}$'' has the same
probability. The full tree of $H$ is used only in the proof.
\end{proof}

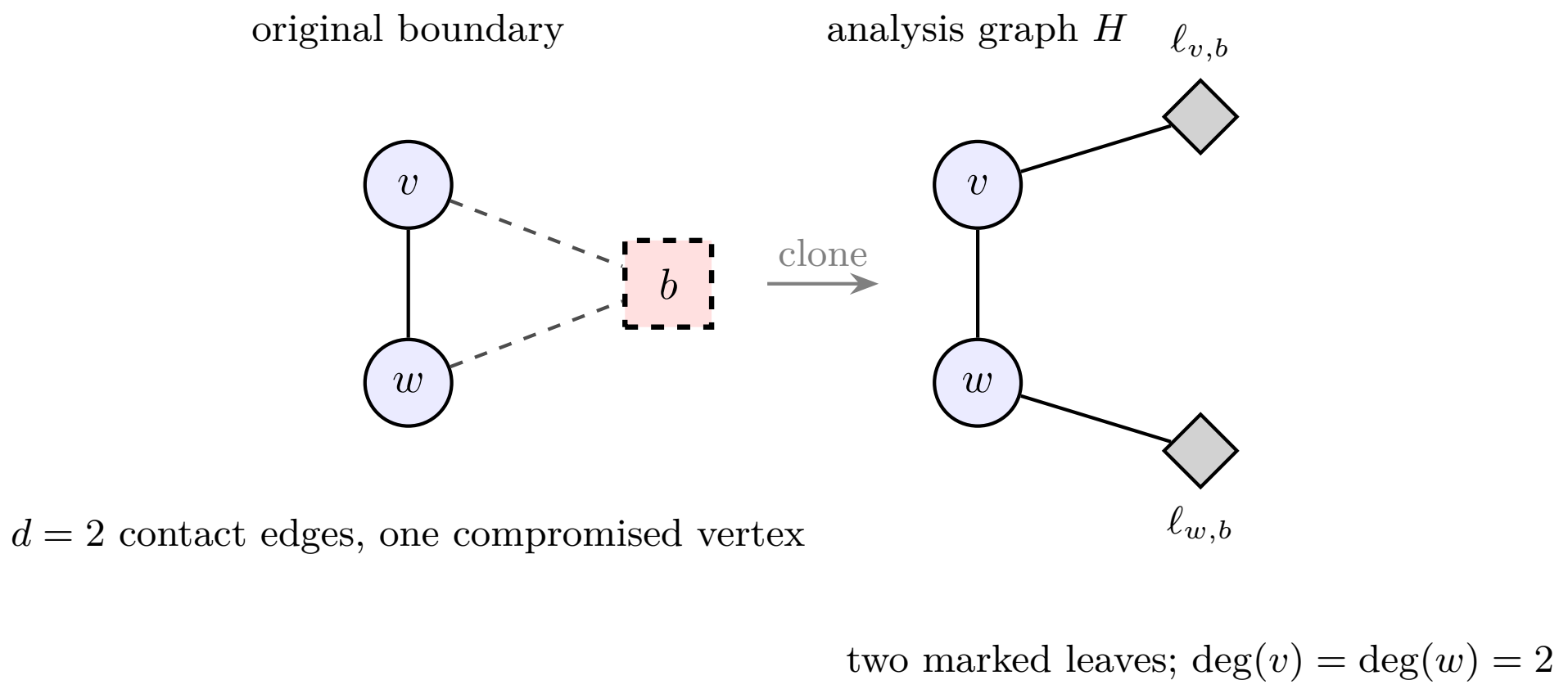
\begin{figure}[H]
\centering
\begin{tikzpicture}[>=Stealth, font=\small]
  \node[honest] (v1) at (0,0.8) {$v$};
  \node[honest] (w1) at (0,-0.8) {$w$};
  \node[cartel] (b) at (2.1,0) {$b$};
  \draw[tree edge] (v1) -- (w1);
  \draw[fatal edge] (v1) -- (b);
  \draw[fatal edge] (w1) -- (b);
  \node[tinylabel, above=6mm of v1] {original boundary};
  \node[tinylabel, below=6mm of w1] {$d=2$ contact edges, one compromised vertex};

  \node[honest] (v2) at (4.6,0.8) {$v$};
  \node[honest] (w2) at (4.6,-0.8) {$w$};
  \node[marked leaf, label={[tinylabel]above:$\ell_{v,b}$}] (lv) at (6.4,1.35) {};
  \node[marked leaf, label={[tinylabel]below:$\ell_{w,b}$}] (lw) at (6.4,-1.35) {};
  \draw[tree edge] (v2) -- (w2);
  \draw[tree edge] (v2) -- (lv);
  \draw[tree edge] (w2) -- (lw);
  \node[tinylabel, above=6mm of v2] {analysis graph $H$};
  \node[tinylabel, below=11mm of lw] {two marked leaves; $\deg(v)=\deg(w)=2$};

  \draw[->, thick, gray] (2.9,0) -- node[tinylabel, above] {clone} (3.8,0);
\end{tikzpicture}
\caption{Cloning on a local boundary. Left: honest $v,w$ joined by an edge, with one cartel vertex $b$ adjacent to both. Right: the analysis graph $H$ keeps $v,w,vw$ but replaces $b$ by two distinct marked leaves $\ell_{v,b},\ell_{w,b}$, preserving $\deg(v)=\deg(w)=2$ and giving $d=2$ marked leaves.}
\label{fig:cloning}
\end{figure}

Figure~\ref{fig:cloning} shows the replacement. Although the new leaves have
different labels, the search still scans corresponding random orders and skips
exactly the same visited vertices until it stops. This is why the replacement
preserves the success probability of this rule.

\subsection{Returning to the target-first rule}\label{app:mainproof}

\begin{lemma}[Plain bound]\label{lem:plain-bound}
For every admissible pair $(G,C)$ with maximum degree at most $\Delta$ and
$d=|E(C,K)|$,
\begin{equation}\label{app:plain}
p_{\mathrm{plain}}(G,C)\ge\frac{1}{\Delta(\Delta-1)^{d-1}}.
\end{equation}
\end{lemma}

\begin{proof}
Apply Lemma~\ref{lem:leaf-bound} and Corollary~\ref{cor:leaf-average} to $H$,
whose maximum degree is at most $\Delta$ and whose marked-leaf set has size $d$,
and use \eqref{app:couple}.
\end{proof}

\begin{lemma}[Taking a revealed target can only help]\label{lem:domination}
For every instance, on the same local permutations,
\begin{equation}\label{app:dom}
\pDFS(G,C)\ge p_{\mathrm{plain}}(G,C).
\end{equation}
\end{lemma}

\begin{proof}
The plain DFS never uses the global override. On any outcome where the plain DFS
succeeds, it visits $t$ before any forbidden vertex. Up to the moment $t$ becomes
visible, the target-first rule follows the same local order and visited-set rule;
formally, by induction on steps, the two runs have the same visited set and the
same scanned prefix at every vertex until the first time $t$ is in the frontier.
At that time the target-first rule takes $t$ and succeeds, possibly earlier than
the plain rule. Hence every plain success is a target-first success.
\end{proof}

\begin{proof}[Proof of Theorem~\ref{thm:degree}]
Lemma~\ref{lem:plain-bound} and Lemma~\ref{lem:domination} give
$\pDFS(G,C)\ge1/[\Delta(\Delta-1)^{d-1}]$, the first inequality of
\eqref{degree:eq-W}. Finally, $d=|E(C,K)|$ is at most the sum of the degrees of the
$c$ vertices of $C$, hence at most $\Delta c$. Since $\Delta-1\ge1$, the map
$x\mapsto1/[\Delta(\Delta-1)^{x-1}]$ is non-increasing, so
$1/[\Delta(\Delta-1)^{d-1}]\ge1/[\Delta(\Delta-1)^{\Delta c-1}]$, the second
inequality. The bound is independent of $n$.
\end{proof}

\subsection{Completion of the matching-example proof}\label{app:sharp}

The main text verifies the count, degrees, admissibility, $d=\Delta c$, and the
probability. It remains to draw the graph without crossings and extend it to
larger sizes.

Draw the path $x_0,\dots,x_{\ell_s-1},q,t$ along a line. Put each $b_l$ above its
block of $\Delta$ consecutive path vertices and join it to them in a fan. The
blocks occupy disjoint intervals, so these edges do not cross. Around each fan
edge $x_ib_{\lfloor i/\Delta\rfloor}$ choose a narrow corridor, disjoint from the
others except at common endpoints. Inside it, replace the edge by $m_i$ parallel
two-edge paths $x_iz_{i,j}w_i$, followed by the edge
$w_ib_{\lfloor i/\Delta\rfloor}$. This network is planar with both endpoints on
its outer face, so it fits inside the corridor. Keeping the original order of corridors at each shared
$b_l$ introduces no crossings there either. All $z,w$ vertices are fresh, so the
result is simple and planar.

For padding, append a chain of honest vertices after the old target $t$ and
declare its last vertex the new target $t'$. The old $t$ and each new internal
chain vertex have degree $2$, and $t'$ has degree $1$. The chain is attached only
at the old target, so no new edge is visible before it is reached. After $q$ the
old target is reached with probability $1$; then every chain vertex has a visited
parent and one unvisited neighbour, so the run is forced to $t'$ with no new
hazard. The probability is unchanged, and every $n\ge n_0$ attains the same
value. The infimum statement follows from \eqref{degree:eq-W} and equality at each
fixed $n\ge n_0$; the example is planar, so the infimum is attained on the planar
subclass. This proves Theorem~\ref{thm:sharp}.

\subsection{Optimality within strict DFS}\label{app:strict-proof}

We prove Theorem~\ref{thm:strict-dfs}. The lower bound is immediate: \RDFS{} uses
the strict stack discipline, its empty-stack fallback is unreachable on an
admissible instance, and Theorem~\ref{thm:degree} gives
$\pDFS(G,C)\ge W(\Delta,c)$ for every admissible $(G,C)$.

For the upper bound fix $A\in\mathcal{D}$ and $n\ge n_0$, put $\ell_s=\Delta c$,
and use the construction of Theorem~\ref{thm:sharp}. Let $p=n-n_0$ and let
$u_0$ be the old target $t$. If $p>0$, append vertices $u_1,\dots,u_p$ with edges
$u_{k-1}u_k$, and set $\tilde t=u_p$; if $p=0$, set $\tilde t=u_0$. Define
$\operatorname{pred}(\tilde t)=q$ when $p=0$, and
$\operatorname{pred}(\tilde t)=u_{p-1}$ when $p>0$. Then $\tilde t$ is adjacent
only to $\operatorname{pred}(\tilde t)$. Label the source $x_0$ by $0$ and
$\tilde t$ by $n-1$, and relabel all other roles uniformly at random. Write $\phi$
for this bijection from roles to labels. The policy may know this distribution,
but not the sampled instance. It is a distribution on
$\mathcal{I}(n,c,\Delta)$, supported on planar instances. Before $q$ is visited
none of $u_0,\dots,u_p$ is visited, and no $x_i$ is adjacent to $\tilde t$.

Fix one realisation of the policy's internal random choices, so $A$ is
deterministic, and let $B_0$ be the sure event. For $0\le i<\ell_s$, let $F_i$
be the event
that the next first visit after $B_i$ is the forward role ($x_{i+1}$ for
$i<\ell_s-1$, and $q$ for $i=\ell_s-1$), and put $B_{i+1}=B_i\cap F_i$. Thus $B_i$ is
the successful path prefix ending at $x_i$ for $i<\ell_s$, and $B_{\ell_s}$ is the
prefix ending at $q$. On $B_i$, record the history immediately after the first
arrival at $x_i$ and revelation of its row, before any further move. Let
$E_i(\tau)$ be the event that this history is $\tau$; these events partition
$B_i$. Let $U_i$ be the new labels in the row at $x_i$: the forward role
and the arms $z_{i,j}$. Then $|U_0|=\Delta$ and $|U_i|=\Delta-1$ for
$1\le i<\ell_s$, independent of $\tau$.

These labels have not appeared before. They represent $x_{i+1}$ (or $q$) and
the $z_{i,j}$, none adjacent to any earlier $x_j$ with $j<i$. The visited vertices
are exactly $x_0,\dots,x_i$, so no earlier revealed row contains a label of $U_i$.

Fix $i$ and an event $E_i(\tau)$ of positive probability. Permute
the labels of $U_i$ arbitrarily and fix every other label. Neither terminal is a
forward role or an arm, so the permutation fixes $0$ and $n-1$. Thus
$\phi^\pi=\pi\circ\phi$ is
another equally likely labelling. It fixes all earlier row sets, the
current row set, the visited set, the chosen labels, and the action determined by
the fixed internal choices; hence $E_i(\tau)$ is invariant. The forward role is
therefore uniform over $U_i$: given the observed transcript, the labels in $U_i$
are equally likely. After arbitrary internal moves through visited vertices no
row is revealed, and the next first visit is determined by the transcript and the
fixed choices or never occurs. Thus
$\Prob(F_i\mid E_i(\tau))\le1/k_i$, with value $0$ if no first visit occurs, where
$k_0=\Delta$ and $k_i=\Delta-1$. This is conditional, and the bounds at
different levels are combined below by the chain rule.

If the next first visit is an arm $z_{i,j}$, the run enters $z_{i,j}$, then must
first-visit $w_i$ (or abort/stall and fail); the frame $z_{i,j}$ cannot be popped
while $w_i$ is unvisited. Entering another arm $z_{i,j'}$ from $w_i$ reveals only
the already visited neighbours $x_i$ and $w_i$. To make another first visit, the
policy must return to the $w_i$ frame. There are finitely many arms, and the
unsafe $b_{\lfloor i/\Delta\rfloor}$ remains unvisited, so $w_i$ cannot be
popped before capture. The run must therefore enter $b$ fatally, stop, or stall.
The target $\tilde t$ is adjacent only to $\operatorname{pred}(\tilde t)$
and is not visible: in the base case $\operatorname{pred}(\tilde t)=q$ is
unvisited, and in the padded case $\operatorname{pred}(\tilde t)=u_{p-1}$ is
unvisited. Thus a wrong arm precludes success, so success is contained in
$B_{\ell_s}$.

When $\Prob(B_i)>0$, partitioning into positive-probability history events
and summing the conditional bounds gives
$\Prob(B_{i+1}\mid B_i)\le1/k_i$; if $\Prob(B_i)=0$, then
$\Prob(B_{\ell_s})=0$. Multiplying the conditional bounds yields
\[
\Prob(\text{success})\le\Prob(B_{\ell_s})
\le\prod_{i=0}^{\ell_s-1}\frac1{k_i}
=\frac1{\Delta}\Bigl(\frac1{\Delta-1}\Bigr)^{\ell_s-1}=W(\Delta,c).
\]

The labelling is uniform on a finite set $\Phi$ of bijections and is independent
of the internal random choices. For a randomised $A$, write
$1_{\mathrm{success}}(\phi,\omega)$ for its success indicator and $P$ for the law
of its internal random choices. Finite linearity gives
\[
\E_\phi\,p_A(\phi)
=\frac1{|\Phi|}\sum_{\phi\in\Phi}\int 1_{\mathrm{success}}(\phi,\omega)\,dP(\omega)
=\int \frac1{|\Phi|}\sum_{\phi\in\Phi}1_{\mathrm{success}}(\phi,\omega)\,dP(\omega)
\le W(\Delta,c),
\]
using the bound just proved for each fixed $\omega$. Hence some fixed
labelled instance in the support has $p_A\le W(\Delta,c)$, because the minimum
over a finite support is at most the average. Since $A$ was arbitrary, the upper
bound holds for $\mathcal{D}$; the planar support and the lower bound give the
planar statement. This proves Theorem~\ref{thm:strict-dfs}.

\section{Proof of the exact tree value}\label{app:tree-proof}

This appendix proves Theorem~\ref{thm:trees}. Instances are those of
Definition~\ref{def:instance}; the rule knows $n$, $c$, and the tree promise, but
not $G$, not $C$, and not the position of the cartel. Fix $c\ge1$ and $n\ge c+2$,
put $h=n-c-2$, the number of vertices that are neither an endpoint nor in the
cartel, and for $h>0$ write $c=\alpha h+\beta$ with $0\le \beta<h$. Here $L$
denotes the length of the honest path.
Section~\ref{tree:h0} handles $h=0$ and the simplification when $n\ge2c+2$.

\subsection{The honest path and its bad children}\label{app:tree-path}

Fix an admissible tree instance $(G,C)$. Let $P=(v_0,\dots,v_L)$ be the unique
$s$-$t$ path in $G$, with $v_0=s$, $v_L=t$, and $L\ge1$. Root $G$ at $s$ and, for
a vertex $v$, let $\mathrm{ch}(v)$ be the set of its children in this rooting.

\begin{lemma}[Path facts]\label{lem:path-facts}
The path $P$ is disjoint from $C$, all $c$ cartel vertices are off $P$, and
\begin{equation}\label{app:tree-length}
L-1\le h.
\end{equation}
\end{lemma}

\begin{proof}
Every $s$-$t$ path in a tree equals $P$. If some $v_i$ belonged to $C$, then
$G-C$ would contain no $s$-$t$ path, contradicting admissibility. Hence $P$ is
honest, and the $c$ cartel vertices lie among the $n-(L+1)$ vertices outside $P$.
Therefore $c\le n-L-1$, which is \eqref{app:tree-length}.
\end{proof}

For $i=0,\dots,L-2$ define
\begin{equation}\label{app:tree-Di}
D_i=\{\,w\in\mathrm{ch}(v_i): w\ne v_{i+1}
\text{ and the subtree at }w\text{ contains a cartel vertex}\,\},
\qquad d_i=|D_i|.
\end{equation}
The subtrees rooted at the vertices of $D_0\cup\dots\cup D_{L-2}$ are pairwise
disjoint and each contains a cartel vertex, so
\begin{equation}\label{app:tree-sumd}
d_0+\dots+d_{L-2}\le c.
\end{equation}
The range stops at $L-2$: $v_{L-1}$ is the parent of $t$, and target-first takes
$t$ as soon as $v_{L-1}$ is visited, before any child outside the path is probed.
Including it would add a spurious factor. Thus the number of risky path vertices
is at most $L-1\le h$.

\subsection{Lower bound: the DFS product}\label{app:tree-lower}

On trees the only visited neighbour of a first-visited vertex is its parent, so
the stored order of the children is a uniform permutation of $\mathrm{ch}(v)$:
drawing a uniform order of the whole neighbourhood and skipping the visited parent
induces that law. The target-first override is applied throughout, including at
$v_{L-1}$, where $t$ is taken before any child outside the path is probed.

\begin{proof}[Proof of Lemma~\ref{lem:dfs-lower}]
We prove
\begin{equation}\label{app:tree-dfsvalue}
\pDFS(G,C)=\prod_{i=0}^{L-2}\frac{1}{d_i+1}\;\ge\;\psi(n,c).
\end{equation}
For $i=0,\dots,L-2$ let $E_i$ be the event that, in the stored permutation of the
children of $v_i$, the honest continuation $v_{i+1}$ precedes every child in
$D_i$. Conditional on the run reaching $v_i$ and on the whole past, the
permutation at $v_i$ is fresh and uniform, and the distinguished children are the
$d_i+1$ vertices $\{v_{i+1}\}\cup D_i$. Each of them is earliest with the same
probability, so
\begin{equation}\label{app:tree-Ei}
\Prob\bigl[E_i \mid E_0,\dots,E_{i-1}\bigr]=\frac{1}{d_i+1}.
\end{equation}

If $E_i$ holds, the search may first explore subtrees at children outside
$D_i\cup\{v_{i+1}\}$; these contain no cartel and no $t$, are finite, and are
exhausted before $v_{i+1}$ is chosen. If $E_i$ fails, some child of $D_i$ is
chosen first; its subtree is finite, contains a cartel vertex but not $t$, so no
override can fire in it and the run is fatal before returning. Thus, if all $E_i$
hold, the search reaches $v_{L-1}$ with no cartel entry and takes $t$ from the
frontier. Success is exactly the intersection of the events, and
\eqref{app:tree-Ei} gives
\begin{equation}\label{app:tree-prod}
\pDFS(G,C)=\prod_{i=0}^{L-2}\frac{1}{d_i+1}.
\end{equation}
For $L=1$ the product is empty and equals $1$; in that case $s$ is adjacent to
$t$, and the target is taken on the first visit.

It remains to compare \eqref{app:tree-prod} with $\psi$. If $h=0$, then
$L-1\le h=0$ forces $L=1$, and the empty product equals $1=\psi(n,c)$. Assume
$h\ge1$ and pad $d_0,\dots,d_{L-2}$ with $h-(L-1)$ zeros. By
\eqref{app:tree-length} the number of zeros is nonnegative, and by
\eqref{app:tree-sumd} the padded list has $h$ nonnegative entries with sum at most
$c$. Lemma~\ref{lem:balanced} bounds the product of the factors $x_i+1$, and
taking reciprocals proves \eqref{app:tree-dfsvalue}.
\end{proof}

\begin{proof}[Proof of Lemma~\ref{lem:balanced}]
Let $h\ge1$ and $c=\alpha h+\beta$ with $0\le \beta<h$. For any nonnegative
integers $x_1,\dots,x_h$ with $x_1+\dots+x_h\le c$ we prove
\begin{equation}\label{app:tree-balancedineq}
\prod_{i=1}^{h}(x_i+1)\le(\alpha+1)^{h-\beta}(\alpha+2)^{\beta}.
\end{equation}
If the sum is below $c$, increasing one entry by $1$ multiplies the product by
$(x_i+2)/(x_i+1)>1$. Hence a maximum over the constraint is attained with sum
exactly $c$.

Suppose two entries satisfy $x_i\ge x_j+2$. Replace $(x_i,x_j)$ by
$(x_i-1,x_j+1)$. The product changes by the factor
\begin{equation}\label{app:tree-factor}
\frac{x_i(x_j+2)}{(x_i+1)(x_j+1)},
\end{equation}
and $x_i(x_j+2)\ge(x_i+1)(x_j+1)$ is equivalent to $x_i\ge x_j+1$, which holds
by assumption. Hence the product does not decrease, and it stays the same or
grows while the sum is preserved. The sum of squares decreases by
$2(x_i-x_j-1)>0$, so this balancing cannot continue indefinitely. It terminates
at a vector with all entries in $\{\alpha,\alpha+1\}$ and sum $c$: exactly
$\beta$ entries equal $\alpha+1$ and $h-\beta$ entries equal $\alpha$. The
product there is the right-hand side
of \eqref{app:tree-balancedineq}. Every vector with sum at most $c$ has product
at most that value, because from any vector one can first raise the sum to $c$
without decreasing the product and then balance without decreasing it.
\end{proof}

Applying Lemma~\ref{lem:balanced} to the padded list gives
\begin{equation}\label{app:tree-padded}
\prod_{i=0}^{L-2}(d_i+1)\le(\alpha+1)^{h-\beta}(\alpha+2)^{\beta},
\end{equation}
and taking reciprocals in \eqref{app:tree-prod} proves
$\pDFS(G,C)\ge\psi(n,c)$ on every tree instance, hence
$\Vtree(n,c)\ge\psi(n,c)$.

\subsection{Upper bound: a distribution of combs}\label{app:tree-upper}

Fix $h\ge1$ and $c=\alpha h+\beta$ with $0\le \beta<h$. Fix any partition of the
$n-2$ non-endpoint labels $\{1,\dots,n-2\}$ into sets
\begin{equation}\label{app:tree-layers}
\Lambda_1,\dots,\Lambda_h,\qquad
|\Lambda_i|=
\begin{cases}
\alpha+2, & 1\le i\le \beta,\\
\alpha+1, & \beta<i\le h.
\end{cases}
\end{equation}
Such a partition exists since $\beta(\alpha+2)+(h-\beta)(\alpha+1)=c+h=n-2$.

Independently for each $i$, sample a uniform label $a_i\in\Lambda_i$. Set
\begin{equation}\label{app:tree-Cprior}
C=\bigcup_{i=1}^{h}\bigl(\Lambda_i\setminus\{a_i\}\bigr),
\end{equation}
and let $G$ have edge set
\begin{equation}\label{app:tree-gpior}
\{\{s,x\}: x\in\Lambda_1\}
\ \cup\ \{\{a_i,x\}: 1\le i<h,\ x\in\Lambda_{i+1}\}
\ \cup\ \{\{a_h,t\}\}.
\end{equation}
The graph is a tree: it has $|\Lambda_1|+\dots+|\Lambda_h|+1=n-1$ edges and is
connected through $s,a_1,\dots,a_h,t$, which is its unique $s$-$t$ path.
A vertex $x\in\Lambda_i\setminus\{a_i\}$ is adjacent only to $a_{i-1}$ (or to $s$
for $i=1$), so it is a leaf; these are exactly the vertices outside that path.
Hence $C$ has size $\sum_i(|\Lambda_i|-1)=c$ and $(G,C)$ is admissible for every
outcome. Let $\mu_{\mathrm{tree}}$ be this distribution on pairs. The pair is
sampled once, before any internal random choice of the rule.

\begin{proof}[Proof of Proposition~\ref{prop:comb}]
Fix a rule $A$ and any realisation $X$ of its internal random source; then $A$ is
deterministic. Write $a_0=s$ and let $H_0$ be the start at $s$. For
$i=1,\dots,h$, let $H_i$ be the event that the run reaches $a_i$ without having
entered a cartel vertex, and stop the analysis immediately after $a_i$ is first
visited, before the next choice. Every vertex outside the unique $s$-$t$ path is
in $C$. On $H_i$, the sequence of first visits is therefore
$s,a_1,\dots,a_i$; safe revisits may occur between them. A first visit outside
this path would be a capture.

At $H_0$ the source row is the fixed set $\Lambda_1$. Let $u$ be the label of the
next first-visited vertex after $H_0$, after any safe revisits; if there is none,
success is $0$. This $u$ is a fixed element of $\Lambda_1$ determined by $X$,
independent of the uniform $a_1$; $H_1$ requires $u=a_1$, hence
\begin{equation}\label{app:tree-H1}
\Prob[H_1\mid H_0]\le \frac{1}{|\Lambda_1|}.
\end{equation}

For $0\le i<h$, suppose $H_i$ occurs. Let $u$ be the next first-visited vertex
after $H_i$, after any safe revisits; if it never occurs, success is $0$. The
entry history contains the observed rows of $s,a_1,\dots,a_i$; in particular the
row of $a_i$ reveals $\Lambda_{i+1}$ as a set. For fixed $X$ and past roles, $u$
is a deterministic function of this information, hence independent of $a_{i+1}$.
Given these observations, every label in $\Lambda_{i+1}$ is still equally likely
to be $a_{i+1}$. The action may use the whole history, the global frontier, $n,c$
and $X$, but these do not identify the next layer's safe vertex.

The frontier at $H_i$ is known explicitly. The visited set is
$\{s,a_1,\dots,a_i\}$, so
\begin{equation}\label{app:tree-frontier}
F(S)=\Bigl(\bigcup_{j=1}^{i}\bigl(\Lambda_j\setminus\{a_j\}\bigr)\Bigr)
\cup\Lambda_{i+1}.
\end{equation}
Every earlier unvisited label is adjacent to $a_{j-1}$, every label of
$\Lambda_{i+1}$ to $a_i$, later path vertices are not adjacent to $S$, and $t$
only to the unvisited $a_h$. Now if $u\notin\Lambda_{i+1}$,
then $u$ is an unvisited cartel leaf of an earlier layer, and choosing it fails.
If $u\in\Lambda_{i+1}$, then $H_{i+1}$ occurs exactly when $a_{i+1}=u$;
otherwise $u$ is a cartel leaf and the run fails. Since $a_{i+1}$ is uniform on
$\Lambda_{i+1}$ and independent of $u$,
\begin{equation}\label{app:tree-chainstep}
\Prob[H_{i+1}\mid H_i]\le\frac{1}{|\Lambda_{i+1}|},
\qquad 0\le i<h.
\end{equation}
The display covers $|\Lambda_{i+1}|=1$, where the only label is $a_{i+1}$ and
the bound is $1$. It is only needed on branches with $\Prob[H_i]>0$.

Chaining the first-hit bounds,
\begin{equation}\label{app:tree-chain}
\Prob_{\mu_{\mathrm{tree}}}[\text{success}]
\le \Prob_{\mu_{\mathrm{tree}}}[H_h]
\le \prod_{i=1}^{h}\frac{1}{|\Lambda_i|}
=\frac{1}{(\alpha+1)^{h-\beta}(\alpha+2)^{\beta}}=\psi(n,c).
\end{equation}
After $H_h$ the target is visible at $a_h$; taking it succeeds and any other
first visit enters the cartel, so the final conditional success probability is at
most $1$. The extra safe vertex $a_h$ prevents the last choice from being bypassed:
the target is not revealed until $a_h$ is reached.

The label prior is finite, so averaging the pointwise bound over $X$ and the
finite label support is a finite sum of measurable success probabilities:
$\E_{\mu_{\mathrm{tree}}}\,p_A(G,C)\le\psi(n,c)$. The minimum over the support is
at most the average, so some fixed pair satisfies $p_A(G,C)\le\psi(n,c)$. Since
$A$ is arbitrary,
\begin{equation}\label{app:tree-upperbound}
\Vtree(n,c)\le\psi(n,c).
\end{equation}
\end{proof}

This proves the upper bound for $h\ge1$. Combined with
Lemma~\ref{lem:dfs-lower}, it gives $\Vtree(n,c)=\psi(n,c)$ in that range.

\section{Secret sharing}\label{sec:bridge-shares}

The routing game has a direct interpretation in cryptographic secret sharing. A
source encodes a key into shares of key material, sends each share in a separate
packet, and the destination reconstructs the key once enough shares arrive.
Assume authenticated links with confidentiality against external eavesdropping.
The adversary reads shares at compromised relays but does not alter messages.
Routing decisions use only observed topology and random choices independent of
the shares; routes, packet lengths and control records reveal no additional
share contents. One game run then tracks one fragment: reaching $t$ before $C$ means
that it arrives unexposed. The game stops at the first exposure because the event
of arriving unexposed has failed, even if a real compromised relay continues
forwarding the fragment.

For XOR reconstruction, fix $M\ge2$ and let the key $\kappa$ take values in
$\{0,1\}^\lambda$, with any distribution. Draw
$X_1,\dots,X_{M-1}$ independently and uniformly from $\{0,1\}^\lambda$,
independently of $\kappa$ and of routing, and set
$X_M=\kappa\oplus X_1\oplus\cdots\oplus X_{M-1}$. The destination recovers
$\kappa=\bigoplus_{i=1}^{M}X_i$. For a fixed proper subset
$I\subsetneq\{1,\dots,M\}$, any specified share tuple and any fixed $\kappa$
have exactly $2^{\lambda(M-1-|I|)}$ consistent mask vectors. Thus the observed
tuple is uniform and independent of $\kappa$: this is perfect privacy, and the
argument is the one-time pad \citep{shannon1949communication}. The receiver needs
all $M$ shares, whereas an adversary given any proper subset learns nothing about
the key. A lost share therefore prevents XOR reconstruction even when privacy is
preserved.

The independent-walker calculation has a precise privacy consequence when each
walker carries a different indexed share, rather than another copy of the same
share. The following statement uses ideal confidential links and passive
observation, as above.

\begin{proposition}[Independent routing of XOR shares]\label{prop:xor-exposure}
Fix an admissible pair $(G,C)$ and the XOR sharing above. Send each of the $M$
indexed shares once using the same policy $A$, with packets stopping at $t$.
Assume their route histories are mutually independent and jointly independent
of the key and masks. Let $I$ be the set of indices whose packets visit $C$
before $t$. Suppose the adversary's view consists only of $I$, the shares
$(X_i)_{i\in I}$, and metadata $Z$ determined by the route histories.
If $p_A(G,C)\ge p_0>0$, then
\begin{equation}\label{eq:xor-exposure}
\Prob(I=\{1,\dots,M\})\le(1-p_0)^M.
\end{equation}
Conditional on any proper exposed set of positive probability, the view is
independent of the key. Moreover, if $P_k$ denotes the view distribution when
the key is set to $k$, then for any $k,k'\in\{0,1\}^\lambda$,
\[
\operatorname{SD}(P_k,P_{k'})\le(1-p_0)^M,
\]
where $\operatorname{SD}$ is total variation distance, the largest difference
in probability assigned to the same event by two distributions.
\end{proposition}

\begin{proof}
Let $E_i$ be the event that packet $i$ visits $C$ before $t$. An unexposed
arrival excludes $E_i$, so $\Prob(E_i)\le1-p_A(G,C)\le1-p_0$. Each $E_i$ depends
only on its own route history. Independence therefore gives
$\Prob(I=\{1,\dots,M\})=\prod_i\Prob(E_i)\le(1-p_0)^M$.
The pair $(Z,I)$ is independent of the key and all shares. For any proper fixed
$J$, the uniform-share calculation above thus remains valid after conditioning
on $I=J$ and on the metadata. Consequently the view laws for any two keys agree
on every proper exposed set. They can differ only on the full-exposure event,
whose probability is the same for both keys and at most $(1-p_0)^M$. This bounds
their total variation distance.
\end{proof}

The route-history assumption is realised by independent executions with their
own memory and fresh random choices, without shared seeds, cross-run adaptation
or a rule that stops one run because another stops. Overlapping routes are
allowed. Under these assumptions the counts in Table~\ref{tab:walker-counts}
also bound full XOR-share exposure by $\varepsilon$. This is a privacy bound;
XOR reconstruction still requires all $M$ shares to arrive. Sending several
copies of one share does not satisfy the one-packet-per-share assumption.

A threshold variant can tolerate lost fragments. Fix $1\le\tau\le M$ and a prime
power $q>M$, encode $\kappa$ in $\mathbb{F}_q$ (a bit string is handled in
blocks), choose $a_1,\dots,a_{\tau-1}$ uniformly and independently from
$\mathbb{F}_q$, independently of $\kappa$ and routing, and set
$f(x)=\kappa+a_1x+\cdots+a_{\tau-1}x^{\tau-1}$, of degree at most $\tau-1$.
Shares are $(x_i,f(x_i))$ at fixed, distinct nonzero $x_i$. Any $\tau$ authentic
shares determine $f$ and $\kappa=f(0)$. For any fixed $r<\tau$ observed shares,
each candidate key has exactly $q^{\tau-1-r}$ consistent coefficient vectors; the
observed values are therefore independent of $\kappa$
\citep{shamir1979how}. This gives perfect privacy for an arbitrary secret
distribution, while allowing the receiver to reconstruct from any $\tau$
authentic shares.

Privacy amplification applies after both endpoints agree on the same raw material
$X$. Let the classical variable $E$ collect the adversary's pre-seed knowledge:
public reconciliation metadata, compromised share contents, and routing or
selection information. A fresh public seed $S$ is uniform and independent of
$(X,E)$, and $h_S$ maps the material to $\lambda$ bits using a 2-universal hash
family: for distinct inputs $x\ne x'$, $\Prob[h_S(x)=h_S(x')]\le2^{-\lambda}$.
The leftover hash lemma states that, if
$\widetilde{H}_{\infty}(X\mid E)\ge k$, then
\[
\operatorname{SD}\bigl((h_S(X),S,E),(U_\lambda,S,E)\bigr)
\le \tfrac12\sqrt{2^{\lambda-k}},
\]
where $U_\lambda$ is uniform and independent of $(S,E)$
\citep[Lemma~2.4]{dodis2008fuzzy}. Average conditional min-entropy is minus the
base-two logarithm of the average best-guess probability
$\max_x\Pr[X=x\mid E=e]$ over $e$. It measures the unpredictability left in $X$.
Choosing $\lambda$ sufficiently below $k$ makes the error small: the derived key
is close to uniform and independent of the public seed and the adversary's
knowledge, even for an unbounded observer. Hashing does not recover missing
material.

The graph parameter $c$ counts compromised relays, while a sharing threshold
counts shares: one compromised relay may observe many fragments. For threshold
sharing, privacy requires fewer than $\tau$ exposed shares, not merely a proper
subset, so Proposition~\ref{prop:xor-exposure} does not directly give its privacy
bound. A complete protocol must also establish enough authentic arrivals and
the entropy available for any extraction step. These requirements remain
separate from a one-fragment routing guarantee.

\section*{Funding}
This work was supported by the project ``Quantum Shield for Data'',
European Commission Recovery and Resilience Facility, under Project
No.~5.1.1.2.i.0/4/25/A/CFLA/005.

\renewcommand{\bibsection}{%
  \section*{\refname}%
  \addcontentsline{toc}{section}{\refname}%
}
\bibliographystyle{hyphennat}
\bibliography{references}

@article{tarjan1972depth,
  author  = {Tarjan, Robert},
  title   = {Depth-First Search and Linear Graph Algorithms},
  journal = {SIAM Journal on Computing},
  volume  = {1},
  number  = {2},
  pages   = {146--160},
  year    = {1972},
  doi     = {10.1137/0201010}
}

@article{papadimitriou1991shortest,
  author  = {Papadimitriou, Christos H. and Yannakakis, Mihalis},
  title   = {Shortest Paths Without a Map},
  journal = {Theoretical Computer Science},
  volume  = {84},
  number  = {1},
  pages   = {127--150},
  year    = {1991},
  doi     = {10.1016/0304-3975(91)90263-2}
}

@article{balamohan2016exploring,
  author  = {Balamohan, B. and Dobrev, S. and Flocchini, P. and Santoro, N.},
  title   = {Exploring an Unknown Dangerous Graph with a Constant Number of Tokens},
  journal = {Theoretical Computer Science},
  volume  = {610},
  pages   = {169--181},
  year    = {2016},
  doi     = {10.1016/j.tcs.2014.07.013}
}

@misc{petrucena2026random,
  author        = {Petru{\v{c}}e\nameSpan{01460061}{\c{n}a}, Kri{\v{s}}j{\=a}nis},
  title         = {Random-Priority Frontier Routing: Tight {$\Theta(n^c)$} Bounds Against {$c$}-Node Cartels},
  year          = {2026},
  eprint        = {2609.00893},
  archivePrefix = {arXiv},
  primaryClass  = {cs.DS},
  doi           = {10.48550/arxiv.2609.00893},
  note          = {arXiv:2609.00893v1}
}

@article{petrucena2026topology,
  author  = {Petru{\v{c}}e\nameSpan{01460061}{\c{n}a}, Kri{\v{s}}j{\=a}nis and Kozlovi{\v{c}}s, Sergejs and V{\=i}ksna, Juris and Kalni\nameSpan{01460061}{\c{n}a}, El{\=i}na and Isaks, Reinis and Celms, Edgars and L{\=a}ce, Lelde and Rencis, Edgars},
  title   = {Topology-Oblivious Random-Walk Key Relaying in Quantum Key Distribution Networks},
  journal = {Entropy},
  volume  = {28},
  number  = {6},
  pages   = {696},
  year    = {2026},
  month   = jun,
  doi     = {10.3390/e28060696}
}

@article{gessel1995tutte,
  author  = {Gessel, Ira M. and Sagan, Bruce E.},
  title   = {The {Tutte} polynomial of a graph, depth-first search, and simplicial complex partitions},
  journal = {The Electronic Journal of Combinatorics},
  volume  = {3},
  number  = {2},
  year    = {1996},
  doi     = {10.37236/1267},
}

@article{shannon1949communication,
  author  = {Shannon, C. E.},
  title   = {Communication Theory of Secrecy Systems},
  journal = {Bell System Technical Journal},
  volume  = {28},
  number  = {4},
  pages   = {656--715},
  year    = {1949},
  doi     = {10.1002/j.1538-7305.1949.tb00928.x}
}

@article{shamir1979how,
  author  = {Shamir, Adi},
  title   = {How to share a secret},
  journal = {Communications of the ACM},
  volume  = {22},
  number  = {11},
  pages   = {612--613},
  year    = {1979},
  doi     = {10.1145/359168.359176}
}

@article{dodis2008fuzzy,
  author  = {Dodis, Yevgeniy and Ostrovsky, Rafail and Reyzin, Leonid and Smith, Adam},
  title   = {Fuzzy Extractors: How to Generate Strong Keys from Biometrics and Other Noisy Data},
  journal = {SIAM Journal on Computing},
  volume  = {38},
  number  = {1},
  pages   = {97--139},
  year    = {2008},
  doi     = {10.1137/060651380},
  note    = {arXiv:cs/0602007}
}

@misc{rfc2104,
  author       = {Krawczyk, Hugo and Bellare, Mihir and Canetti, Ran},
  title        = {{HMAC}: Keyed-Hashing for Message Authentication},
  howpublished = {RFC 2104},
  year         = {1997},
  doi          = {10.17487/RFC2104}
}

@inproceedings{karp2000gpsr,
  author    = {Karp, Brad and Kung, H. T.},
  title     = {{GPSR}: Greedy Perimeter Stateless Routing for Wireless Networks},
  booktitle = {Proceedings of the 6th Annual International Conference on Mobile Computing and Networking},
  pages     = {243--254},
  year      = {2000},
  doi       = {10.1145/345910.345953}
}

@article{deng1999exploring,
  author  = {Deng, Xiaotie and Papadimitriou, Christos H.},
  title   = {Exploring an unknown graph},
  journal = {Journal of Graph Theory},
  volume  = {32},
  number  = {3},
  pages   = {265--297},
  year    = {1999},
  doi     = {10.1002/(SICI)1097-0118(199911)32:3<265::AID-JGT6>3.0.CO;2-8}
}

@inproceedings{barnoy1991canadian,
  author    = {Bar-Noy, Amotz and Schieber, Baruch},
  title     = {The {Canadian} Traveller Problem},
  booktitle = {Proceedings of the Second Annual ACM-SIAM Symposium on Discrete Algorithms},
  pages     = {261--270},
  publisher = {SIAM},
  year      = {1991},
  url       = {https://dl.acm.org/doi/10.5555/127787.127835}
}

@article{dobrev2006searching,
  author  = {Dobrev, Stefan and Flocchini, Paola and Prencipe, Giuseppe and Santoro, Nicola},
  title   = {Searching for a black hole in arbitrary networks: optimal mobile agents protocols},
  journal = {Distributed Computing},
  volume  = {19},
  number  = {1},
  year    = {2006},
  doi     = {10.1007/s00446-006-0154-y}
}

@inproceedings{yao1977probabilistic,
  author    = {Yao, Andrew C.-C.},
  title     = {Probabilistic computations: Toward a unified measure of complexity},
  booktitle = {18th Annual Symposium on Foundations of Computer Science},
  pages     = {222--227},
  publisher = {IEEE},
  year      = {1977},
  doi       = {10.1109/SFCS.1977.24}
}

@article{dolev1993perfectly,
  author  = {Dolev, Danny and Dwork, Cynthia and Waarts, Orli and Yung, Moti},
  title   = {Perfectly secure message transmission},
  journal = {Journal of the ACM},
  volume  = {40},
  number  = {1},
  pages   = {17--47},
  year    = {1993},
  doi     = {10.1145/138027.138036}
}

@article{salvail2010security,
  author  = {Salvail, Louis and Peev, Momtchil and Diamanti, Eleni and All{\'e}aume, Romain and L{\"u}tkenhaus, Norbert and L{\"a}nger, Thomas},
  title   = {Security of trusted repeater quantum key distribution networks},
  journal = {Journal of Computer Security},
  volume  = {18},
  number  = {1},
  pages   = {61--87},
  year    = {2010},
  doi     = {10.3233/JCS-2010-0373}
}

@inproceedings{childs2003exponential,
  author    = {Childs, Andrew M. and Cleve, Richard and Deotto, Enrico and Farhi, Edward and Gutmann, Sam and Spielman, Daniel A.},
  title     = {Exponential algorithmic speedup by a quantum walk},
  booktitle = {Proceedings of the 35th Annual ACM Symposium on Theory of Computing},
  pages     = {59--68},
  publisher = {ACM},
  year      = {2003},
  doi       = {10.1145/780542.780552}
}

\end{document}